\documentclass{aa}
\usepackage{amsmath}
\usepackage{psfig}
\usepackage{astrobib}
\usepackage{journals}

\def\aap{A\&A}
\def\apj{ApJ}
\def\apjl{ApJL}
\def\mnras{MNRAS}

\def\lesssim{\mathrel{\hbox{\rlap{\hbox{\lower4pt\hbox{$\sim$}}}\hbox{$<$}}}}
\def\gtrsim{\mathrel{\hbox{\rlap{\hbox{\lower4pt\hbox{$\sim$}}}\hbox{$>$}}}}

\newcommand\schodel{Sch\"odel~}

\def\msun{{\,M_\odot}}
\def\rsun{{\,R_\odot}}
\def\lsun{{\,L_\odot}}

\newcommand\fe{Fe K$\alpha$\ }

\newcommand\tstar{T_*}
\newcommand\nd{n_{\rm d}}

\newcommand\sgra{Sgr~A$^{*}$ }

\newcommand\td{T_{\rm d}}
\newcommand\lmax{L_{\rm max}}
\newcommand\tchar{T_{\rm char}}

\def\>{$>$}
\def\<{$<$}

\def\simlt{\lower.5ex\hbox{$\; \buildrel < \over \sim \;$}}
\def\simgt{\lower.5ex\hbox{$\; \buildrel > \over \sim \;$}}
\def\sqr#1#2{{\vcenter{\hrule height.#2pt
      \hbox{\vrule width.#2pt height#1pt \kern#1pt
         \vrule width.#2pt}
      \hrule height.#2pt}}}

\def\del#1{{}}

\begin{document}

\title{X-ray flares from Sgr A$^*$: star-disk interactions?}

\author{Sergei Nayakshin\inst{1} \and Jorge Cuadra\inst{1} \and Rashid
Sunyaev\inst{1,2}} \institute{Max-Planck-Institut f\"ur Astrophysik,
Postfach 1317, D-85741 Garching, Germany\\
\and Space Research Institute, Moscow,
Russia\\}

\titlerunning{X-ray Flares and inactive disk in Sgr A*}
\authorrunning{Nayakshin et al.}
\offprints{serg@mpa-garching.mpg.de}

\date{A\&A, submitted}

\abstract{Sgr A$^*$, the putative black hole in our Galactic Center
(GC), is extraordinary dim in all frequencies. Apparently the black
hole is unable to accrete at the Bondi accretion rate for some
reason. Another mystery of \sgra is the recently discovered large
magnitude and short duration X-ray flares (Baganoff et al. 2001).
Here we point out that X-ray flares should be expected from star
passages through an inactive (i.e. {\em formerly} accreting) disk.
There are thousands of stars in \sgra stellar cluster, and each star
will pass through the disk twice per orbit. A shock hot enough to emit
X-rays is formed around the star when the latter is inside the
disk. We develop a simple yet somewhat detailed model for the X-ray
emission from the shock. The duration of the flares, their X-ray
spectra, frequency of events, and weakness of emission in the radio
and near infra-red appear to be consistent with the available
observations of X-ray flares.  We therefore suggest that at least a
fraction of the observed flares is due to the star-disk passages.
Such star-disk flares are also of interest in the nuclei of nearby
inactive galaxies, especially in connection with perspective
Constellation-X and XEUS X-ray missions. \keywords{Galaxy: center --
X-rays: galaxies -- accretion, accretion disks -- black hole physics }
} \maketitle

\section{Introduction}\label{sec:intro}

\subsection{\sgra: underfed X-ray flaring black hole}\label{sec:underfed}

With the current high quality data (e.g., \schodel et al. 2002, Genzel
et al. 2003; Ghez et al. 2003a,b) it seems very safe to say that the
center of our Galaxy hosts a supermassive black hole, with $M_{BH}
\simeq 3 \times 10^6 \msun$. The black hole is associated with \sgra
source, the compact radio core \cite{ReidReadheadVermeulen1999},
discovered by Balick \& Brown (1974). Thanks to Chandra observations
(Baganoff et al 2003a), we know that there is enough of hot gas at
distances of the order $\sim 10^{17}$ cm (or $\sim$ 1\arcsec) to make
the black hole glow at around $\sim$ few $\times 10^{40}$ erg/sec if
accretion proceeds via the standard radiatively efficient disk
(Shakura \& Sunyaev 1973). However the observed bolometric luminosity
of \sgra is 4 orders of magnitude smaller than expected (e.g. Zhao,
Bower \& Goss 2001; see also review by Melia \& Falcke 2001). The
luminosity in X-rays is even smaller and is $\sim 10^{-7}$ of the
nominally expected numbers (Baganoff et al. 2003a). Understanding of
this discrepancy is the key to accretion physics at low accretion
rates which encompasses a very large scope of objects from galactic
centers to stellar mass compact objects (e.g., Narayan 2002).

The currently leading explanation of the low luminosity from \sgra is
the Non-Radiative Accretion Flows (NRAFs), which were discussed in
great detail first by Narayan \& Yi (1994). The corner stone
assumption of this model -- that ions are much hotter than the
electrons in the inner region of the flow -- has not yet been
independently verified. Initially it was believed that these flows
carry (``advect'') most of their energy into the black hole, radiating
very little on the way. Several years ago it became clear that these
hot flows also drive powerful thermal winds (Blandford \& Begelman
1999), so the accretion rate onto the black hole is significantly
reduced. This theoretical picture seems to be confirmed in its gross
features by the observations (e.g., Narayan, Yi \& Mahadevan 1995;
Bower et al. 2003; Quataert 2003), although alternative
interpretations of \sgra also exist (Melia 1994; Falcke \& Markoff
2000; Coker \& Melia 2000; Yuan, Markoff \& Falcke 2002).

The recently discovered X-ray flares in \sgra (Baganoff et al. 2001)
brought further theoretical challenges. The X-ray flares happen at a
rate of about one per day (Goldwurm et al. 2003a; Baganoff et
al. 2003b). The X-ray luminosity during the flares exceeds the
quiescent one by up to a factor of $100$ (although most of flares are
weaker than that).  Flare durations range from $\sim 1$ to 10
kiloseconds. The X-ray spectrum of flares is harder than the quiescent
one. The follow-up multi-wavelength campaign of \sgra has not detected
significant variability in any other band except for X-rays (Baganoff
et al. 2003b; see also Hornstein et al. 2002). This latter fact is
particularly troubling since in all the current models the X-ray
flares occur at the inner part of the flow that emits \sgra radio
emission. Thus the challenge is to understand why such a large
variation in X-rays does not lead to flares in other frequencies.

\subsection{Cold disks in LLAGN and \sgra (?)}\label{sec:llagn} 

It is believed that accretion flow physics at low accretion rates
should be largely self-similar (e.g. Narayan 2002) and the difference
is just in scales (e.g $M_{\rm BH}$ and accretion rate $\dot{M}$).
Therefore \sgra is expected to be physically similar to the Low
Luminosity AGN (LLAGN; e.g. Ho 1999). Most of LLAGN appear to posses
thin cold disks. The presence of these disks is deduced convincingly
through (i) water maser emission (Miyoshi et al. 1995); (ii) infrared
``bumps'' in the spectral energy distribution of LLAGN (Ho 1999) that
are thought (Quataert et al. 1999) to be the thermal disk emission
(with disk temperatures $T\simlt 10^3$ K); and (iii) the prevalence of
the double-peaked broad emission lines (e.g. Ho 2003).  The outer
radii of these disks are poorly constrained, whereas the inner disk
radii can be constrained from both the SED and the emission line
profiles, and seem to be in the range $\sim 100-1000 R_g$, where $R_g$
is the gravitational radius.

It is thus tempting to suggest that such a disk may also exist in our
GC (see \S \ref{sec:isdisk}). There appears to be several promising
ways to test the disk hypothesis. The simplest of all is through
eclipses of close stars (Nayakshin \& Sunyaev 2003). In particular,
the orbit of star named S2 is very well determined
(\schodel et al. 2002, Ghez et al. 2003b).  Another important effect
is the reprocessing of the optical-UV radiation of the star in the
disk into the near infrared (NIR) band, which can yield a strong
increase in the perceived NIR flux of the star. The current S2 data
alone practically rule out an {\em optically thick} disk (Cuadra,
Nayakshin \& Sunyaev 2003). However a disk optically thin in NIR is
permitted.

\subsection{X-ray flares in \sgra: star disk interactions?}\label{sec:flares}

In this paper we point out an additional effect as a test of the
inactive disk hypothesis. Recent results of Genzel et al. (2003) show
that there are as many as $\sim 10^4$ stars in the inner arcsecond
($\sim 10^5 R_g$) of \sgra. Several of these stars {\em cross} the
inactive disk every day. While inside the disk, the stars drive a very
strong and hot shock wave in the disk material. The shock will emit
X-rays (since the shock is short lived, this emission is actually a
flare). In this paper we shall build a simple yet somewhat detailed
and self-contained model for these events.

Let us point out that such star-disk flares have a potential to
naturally explain why there are no counterparts of X-ray flares in the
radio and near infrared frequencies. A somewhat unexpected feature of
this model is that {\em the flares are the result of a process that
has nothing to do with the hot gas that is responsible for \sgra
quiescent emission in all frequencies.} This is so because stars are
``small'' ($\sim 10^{11}$ cm) compared with the size of the hot flow
($\sim 10^{17}$ cm), and they have a negligible direct effect on the
flow of hot gas. Flares are emitted when the stars strike the cold
disk that is much denser than the hot flow. This happens at radii of
order $10^3-10^5 R_g$, and thus the flares do not interfere with the
flow at $R\simlt 10_g$ where the jet originates. In addition, the
X-ray flares are emitted by a thermal moderately hot gas, and we will
later see that the radio and near infrared flares from the star-disk
passages are weak.

Finally, the problem of X-ray flares from star-disk passages merits a
careful study even if there is no disk in our own Galaxy. Similar
X-ray flares may be detectable in other LLAGN, especially the close
ones such as M31 or M32 (Ho, Terashima \& Ulvestad 2003).

The plan of the paper is as following. We first attempt to determine
in \S \ref{sec:disk} the kind of a disk that may reside in \sgra from
arguments that are independent from the observed X-ray flares. We find
that the disk must be inactive, i.e. not accreting, with temperature
$\simlt$~few~$\times 10^2$ K, with gas density in the range of
$10^{10}$ to $10^{15}$ cm$^{-3}$. In \S \ref{sec:temperature} we will
see that a disk optically thick in near infrared is ruled out but an
optically thin disk is allowed. In \S \ref{sec:cluster} we calculate
the rate or the star-disk passages, and also show that the disk is
very likely to have an inner hole. In \S \ref{sec:single} we show
schematics of a star-disk passage and mention the connection of our
work to the existing literature on such events. X-ray luminosity and
duration of flares is calculated in \S \ref{sec:xlum}. X-ray spectra
are considered in \S \ref{sec:semit}, whereas the near infrared
spectra are addressed in \S \ref{sec:nir}. In \S \ref{sec:typ}
properties of flares produced by a low mass star and a high mass star
are discussed, and in \S \ref{sec:comp} a comparison between the model
and observations is made. A list of shortcomings of our work is given
in \S ref{sec:short}. Finally, discussion of the results is presented
in \S \ref{sec:discussion}.

\section{Constraints on the disk properties}\label{sec:disk}

\subsection{Arguments for and against a disk in Sgr~A$^*$}\label{sec:isdisk}

If the putative accretion disk were of the standard type (Shakura \&
Sunyaev 1973), then it would have been seen by now even if accretion
rate is as small as $10^{-9} \msun$/year (Narayan 2002). But at very
low disk accretion rates hydrogen in the disk becomes neutral and the
accretion rate plunges to very low values (e.g. see review by Cannizzo
1998; see also equation \ref{mdot} below). This is why these disks are
inactive, i.e. not accreting. In addition, these disks seem to miss
their inner regions for a variety of reasons (e.g., Quataert et
al. 1999 and \S \ref{sec:effects}) and therefore they can escape the
near infra-red (NIR) detection limits that rule out standard
steady-state disks.

However, Falcke \& Melia (1997; FM97 hereafter) probed the existence
of the disk through its interaction with the winds from hot
stars. This wind is believed to be the source of the hot gas that
should be accreting onto the black hole. FM97 assumed that the wind
runs into and settles down onto the disk, releasing the mechanical
energy into the black-body radiation on impact. They found that unless
the wind has a large specific angular momentum (so that it cannot
penetrate to small radii), the luminosity expected in the NIR violates
the available limits on \sgra NIR emission. Since there are $\simgt
10$ wind sources, no large angular momentum in the wind was expected.
FM97 therefore stated that existence of the disk in \sgra is highly
unlikely (see also Coker, Melia \& Falcke 1999).

New data show that eleven out of twelve wind-producing stars rotate in
the same (counter-Galaxy) direction (Genzel 2000). These and other hot
luminous stars in the region share the same puzzling property: they
are quite young (Ghez et al. 2003b; Genzel et al. 2003) and it is
unclear how these stars made it to the innermost arcsecond of the
Galaxy. One of the possibilities is that these stars were created from
a single large molecular cloud that fell into the inner region of the
Galaxy with a small impact parameter. This cloud could have created
both the stars and an accretion disk whose (tiny not accreted) remnant
may still be there.\footnote{We have recently learned that, using data
of R. Genzel's group, Levin \& Beloborodov 2003 found that 10 out of
13 wind-producing stars may lie ``exactly'' in a single plane, which
further confirms our suggestions.} Note also that matter in the
accretion disk flows inward and {\em outward} from the characteristic
(circularization) radius (Kolykhalov \& Sunyaev 1980).  It is thus not
possible to accrete all of the disk in the active phase without
leaving a remnant.

Nayakshin (2003) recently showed that, in the setting similar to FM97,
the hot flow rapidly looses its heat to the inactive disk. The latter
re-radiates this energy in the infrared. As the hot gas looses heat,
it also looses vertical pressure support, and as a result condenses
onto the inactive disk.  The accretion flow in the radial direction is
essentially terminated. Thus the hot gas is accreting not on the black
hole but rather on the inactive disk.  The tightest of the existing
NIR limits is not violated by the model if wind circularization radius
is $R_c \simgt 3\times 10^4 R_g$. Since the wind sources are a few
arcsecond ($= \hbox{few}\times 10^5 R_g$) away from \sgra (e.g. Genzel
et al. 2003), it appears that this ``large'' value of the
circularization radius is quite reasonable. The luminosity emitted by
\sgra is very small in this model because the gravitational potential
at the condensation point is tiny compared with that of the last
stable orbit around the black hole. We refer to this model as the
``Frozen Accretion'' (FA) model of Sgr~A$^*$, and use some of the
constraints from the model to limit the possible size and density in
the disk.

\del{Finally note that an inactive accretion disk is very flat, e.g., $H/R
\sim 10^{-3}$ (see text after equation \ref{h}), and hence if the disk
happens to be viewed nearly edge-on, then its thermal emission is
hardly detectable, alleviating arguments of both
\citeN{FalckeMelia1997} and \citeN{Narayan2002} even further.}

\subsection{Disk temperature (and NIR optical depth)}\label{sec:temperature}

First consider a disk optically thick in all frequencies. Let $R_d$ be
the outermost radius of the disk. Since the disk is inactive, the
effective temperature of such a disk does not behave as the usual $\td
\propto R^{-3/4}$. For example, $\td$ appears to stay nearly
independent of radius for quiescent disks of eclipsing dwarf novae
(e.g. Figs. 5 \& 6 in Bobinger et al. 1997). If $\td\simeq$ const,
then the disk total energy output is dominated by the emission from
outermost radii. Accordingly the disk luminosity is $L_d \sim 2 \pi
R_d^2 \sigma_B \td^4$. In order to satisfy observational constraints
on the total luminosity of \sgra in quiescence, $L_{\rm q}$, this
temperature must be smaller than
\begin{equation}
\td < \left[\frac{L_{\rm q}}{2\pi R_d^2 \cos{i} \sigma_B}\right]^{1/4}
= 77 \; {\rm K}\; \frac{L_{36}^{1/4}}{\cos{i}^{1/4} r_4^{1/2}}\;,
\label{td}
\end{equation}
where $L_{36} = L_{\rm q}/10^{36}$ erg/sec, $r_4 = R_d/10^4 R_g$, $R_g
= 2GM/c^2 = 9\times 10^{11}$ cm is the gravitational radius for the
black hole mass of $M_{BH} = 3 \times 10^6 \msun$. We shall see below
that the disk is unlikely to be very optically thick so the disk
midplane temperature should not exceed the effective temperature by a
large number.

We should also require the disk to be hot enough to re-radiate the
visible and UV radiation of the \sgra star cluster itself. We use the
latest results on the star number counts in \sgra cluster from Genzel
et al. 2003, together with the fact that the total luminosity of stars
in the central parsec of the Galaxy is about 50 million Solar
luminosities ($\lsun$). We first calculate the light to mass ratio,
and then use equation \ref{mrs} to find that the integrated light of
the star cluster within radius $R$ is
\begin{equation}
L_*(R) = 2.1\times 10^4 \;\lsun\; r_4^{3/2} \;\hbox{erg/sec}
\label{lsgra}
\end{equation}
The flux incident on the disk is then $F_*(R) = L_*/4\pi R^2$, and the
resulting effective temperature due to stellar radiation heating is
\begin{equation}
\tstar = 190 \;\hbox{K}\; r_4^{-1/8}\;,
\label{tsgra}
\end{equation}
which is a factor of 3 higher than the value in equation \ref{td}.
This latest result shows that an optically thick disk in fact would
over-produce the infrared luminosity of \sgra by about a factor of a
hundred (e.g. see \sgra spectrum in Falcke \& Melia 2001 and in
Narayan 2002).  Therefore the disk must be optically thin in the
infrared frequencies {\em or} highly inclined with respect to our line
of sight. An estimate shows that for not too small disk inclination
angles the infrared optical depth of the disk should be no larger than
about 0.01 at frequency $\sim 10^{13}$ Hz. In addition, close passages
of very bright stars such as S2 ($L\sim 10^5\lsun$; Ghez et al. 2003b)
may actually increase the value of $\tstar$ to about 300 K (Cuadra et
al. 2003).

\subsection{Disk size}\label{sec:size}

The quiescent spectrum of \sgra is in rough accord with the data if
the main part of the hot flow condenses onto the cold disk at radius
$R_c\simgt 3\times 10^4 R_g$ (Nayakshin 2003). If condensation occurs
at much smaller radii, then the flow gains too much gravitational
energy (while falling from the Bondi radius, $R_{\rm B}\sim 10^5 R_g$
to $R_c$) and the luminosity of the disk would be too high. The value
of $R_c$ could be however reduced if the exact accretion rate in the
condensing flow is lower than the Bondi estimate (similar to the ADAF
case; see \S 5.2 in Narayan 2002). Therefore we estimate the minimum
disk outer radius as
\begin{equation}
R_d \sim 1-10 \times 10^4 R_g
\label{rc}
\end{equation}
There are no strong constraints based on the FA model on the inner
disk radius, $R_i$, as long as it is at least a factor of few smaller
than condensation radius $R_c$.

\subsection{Constraints from close bright star passages}\label{sec:size}

The second string of constraints on the disk size and inclination
angle, $i$, comes from the fact that close stars could be eclipsed by
the disk {\em if} it is optically thick in near infrared. Here we
summarize our preliminary results (Cuadra et al. 2003; see also
Nayakshin \& Sunyaev 2003) based on the measured S2 star's
positions. The parameters of the orbit are now very well known
(Sch\"odel et al. 2002; Ghez et al. 2003b). Cuadra et al. (2003) find
that if the disk is optically thick and has an inner hole less than
$\sim 0.01\arcsec\;\sim 10^3 R_g$, then the outer disk radius may not
much exceed $10^4 R_g$, barely making the constraints based on the FA
model. In addition, for an inner hole of $\simeq 0.03$ arcsecond,
there is a large range in the disk orientation parameter space that no
eclipses of S2 are predicted for any arbitrarily large value of the
outer disk radius.

However reprocessing of the optical-UV radiation received by the disk
from the star into the NIR band could significantly increase the
apparent star brightness if the star is close enough to the
disk. Using the deduced S2 orbit, we studied this effect for all
possible disk orientations, assuming an optically thick disk with no
inner hole and infinitely large outer radius. We find (Cuadra et
al. 2003) that an increase in the 2.2$\mu$m flux of S2 should have
been detected at least in one of the S2 measured positions, whereas
observations do not confirm that. Therefore the disk, if any, must be
optically thin in the NIR. We are currently modeling similar
reprocessing for an optically thin disk. Unfortunately these
constraints are not easily converted into constraints on the total
column depth of the disk, $\Sigma$ (see \S \ref{sec:tau}), because the
properties of the dust grains in the disk are very uncertain (see \S
\ref{sec:nir}).

We have also checked some other potentially observable indicators of
the disk presence. Due to star's UV irradiation, a layer of completely
ionized hydrogen forms on the top of the disk. Assuming temperature of
order $10^4$ K for the layer, we found that its column depth is
$N_H\sim 10^{18} n_{11} r_{15}^{-2}$ cm$^{-2}$, where $r_{15}$ is
distance between the star and the disk in units of $10^{15}$ cm, and
$n_{11}$ is the hydrogen nuclei density in units of $10^{11}$
cm$^{-3}$. We then estimated the Brackett $\gamma$ ($\lambda=2.16
\mu$m) line flux from this photo-ionized layer of gas, and compared it
with that of the S2 star in the K-band. This flux is about $10^{-3}$
of the star's continuum $\nu F_{nu}$ at that frequency. For
comparison, this is about 30 times weaker than the absorption in the
Br $\gamma$ line from the star itself (see Fig. 1 in Ghez et
al. 2003b). In addition, the emission line from the photo-ionized
layer should be 5-10 times broader than the line from the star due to
much faster disk rotation, making it challenging to observe.

Finally, free-free absorption in the photo-ionized layer is important
for frequencies smaller than $\sim 10^{11} r_{15}^{-1}$ Hz. For these
frequencies, the corresponding thermal radio emission of the layer is
at the level of about 1 percent of the \sgra jet emission at the
respective frequencies, rendering it negligible.

\subsection{Minimum disk density}\label{sec:nmin}

\paragraph{Pressure equilibrium argument.} In the frozen accretion
model of Nayakshin (2003), the inactive disk and the hot accretion
flow are in direct physical contact. Thus the disk midplane pressure
should be at least as large as the pressure in the hot flow. Assuming
the disk consists mainly of molecular hydrogen, $P_d\simeq k \nd
\td/2$, where $\nd$ is the midplane density of hydrogen nuclei in
cm$^{-3}$. The density in the hot accretion flow can be estimated
through $\dot{M} \sim 4 \pi R^2 v_R m_p n_{\rm hot}$, where $v_R \sim
\alpha \sqrt{GM_{\rm BH}/R}$ and $\alpha = 0.1 \alpha_1$ is the
viscosity parameter (Shakura \& Sunyaev 1973).  This yields
\begin{equation}
\nd \;\simgt \; 1.5 \times 10^{9}
\quad\hbox{cm}^{-3}\quad \frac{\dot{M}' T'}{\alpha_1 T_2} \;
\left(\frac{3\times 10^4 R_g} {R_c}\right)^{3/2}\;,
\label{nmin}
\end{equation}
where $\dot{M}'\equiv \dot{M}/3\times 10^{-6}\msun$ year$^{-1}$ is the
accretion rate in the hot flow, $T' = T_{\rm hot}/3 \times 10^7$ K is
dimensionless temperature of the hot flow, $R_c$ is the wind
circularization radius, and $T_2 = \td/100$ K is the disk unperturbed
midplane temperature.

\paragraph{Total accreted mass argument.} 
If condensation of the hot flow has been going on for a time $t_a$,
then the mass accreted by the disk from the wind is $\Delta M\sim
\dot{M} t_a$. If the accreted mass is deposited over area $\sim \pi
R_c^2$, then the accreted surface density is
\begin{equation}
\Delta \Sigma_d \sim 3 \;\hbox{g/cm}^2\; \dot{M}' t_{a6}
\left(\frac{3\times 10^4 R_g} {R_c}\right)^2\;,
\label{dsigma}
\end{equation}
where $t_{a6}=t_a/10^6$ years. This translates into
the disk density increase of
\begin{equation}
\Delta \nd \sim 4 \times 10^{10} \;\hbox{cm}^{-3}\; \dot{M}' t_{a6}
T_2^{-1/2}\; \left(\frac{3\times 10^4 R_g} {R_c}\right)^{7/2}\;.
\label{dn}
\end{equation}
This estimate yields a larger value for the minimum disk density than
does equation (\ref{nmin}), so we conclude that the minimum disk
density is in the range of $\sim \hbox{few}\times 10^{10}$ hydrogen
nuclei in cm$^{3}$. Note however that the above arguments are
applicable to the disk radii with $R\sim R_c$, and unfortunately they
cannot be robustly extended to radii much smaller than that.

\del{As Nayakshin (2003)
finds that circularization radius $R_c\simgt 3\times 10^4 R_g$, then
the constraint (\ref{nmin}) is applicable only to such large radii and
roughly yields min\{$\nd$\}$\sim 10^9$ cm$^{-3}$.}

\subsection{Disk thickness}\label{sec:thick}

The disk half thickness, $H$, is given by the
hydrostatic balance (we neglect self-gravity; see below): 
\begin{equation}
H = \sqrt{\frac{k T R^3}{GMm_p}}= 3.9 \times 10^{12} \, T_2^{1/2}
r_4^{3/2}\; {\rm cm}\;,
\label{h}
\end{equation}
where we assumed that the gas is dominated by molecular hydrogen so
$\mu\simeq 2 m_p$. Note that the star's radius ($\sim 10^{11}$ cm) is
much smaller than the disk height scale for $R > 10^3 R_g$. The disk
is very flat since $H/R = 0.96 \times 10^{-3} T_{2}^{1/2}\,
r_4^{1/2}$.

\subsection{Mass of the disk and maximum density}\label{sec:mass}

The mass of the accretion disk with a constant (with radius) midplane
density and temperature, is
\begin{equation}
M_d \simeq 0.1 \msun \, n_{11} \left(\frac{R_d}{10^4 R_g}\right)^{7/2}
T_2^{1/2}\;,
\label{md}
\end{equation}
where $n_{11}\equiv \nd/10^{11}$ cm$^{-3}$ is the midplane density of
hydrogen nuclei.  $M_d$ is much smaller than the black hole mass so
that the disk is not globally self-gravitating (i.e., in the radial
direction; Shlosman \& Begelman 1989) and will not {\em
gravitationally} influence stellar dynamics in Sgr~A$^*$. The disk is
locally (vertically) self-gravitating when $Q = M_d/(H/R)M_{BH} \simgt
1$ (Goldreich \& Lynden-Bell 1965; Kolykhalov \& Sunyaev 1980; Gammie
2001).  We have
\begin{equation}
Q = \frac{M_d}{(H/R) M_{BH}} = 5.0 \times 10^{-5} \; r_4^3 n_{11}\;,
\label{q}
\end{equation}
i.e. the disk is not vertically self-gravitating as well for the
nominal values of $\nd$ and $R_d$ in this equation. We can turn the
question around to derive the maximum disk density as the density at
which the disk would become self-gravitating. In the latter case it
would presumably break up into self-gravitating rings or regions and
the star-disk impacts would become too rare. Hence,
\begin{equation}
\nd \;\simlt \; 10^{15} r_4^{-3}
\label{nm}
\end{equation}
For $r_4 = 3$ equation (\ref{nm}) yields maximum density
$\sim$~few~$\times 10^{13}$ cm$^{-3}$. Note that equation \ref{nm} is
applicable for any radius $r_4$ in contrast to equations \ref{nmin}
and \ref{dn}.

\del{One should note that perhaps the disk can be ``slightly''
self-gravitating so that there are regions of the disk dominated by
self-gravity. But most likely these regions would occupy only a small
fraction of the disk surface area and the impacts of stars on these
regions would be rare enough to neglect.}

\subsection{Column depth and accretion rate}\label{sec:tau}

The disk surface density is $\Sigma \simeq 2 H n_d m_p$, and is equal
to
\begin{equation}
\Sigma = 1.3 \; n_{11} r_4^{3/2}\, T_2^{1/2} {\rm g\; cm}^{-2}\;.
\label{sigma}
\end{equation}
It is therefore clear that with the minimum and maximum disk density
values just found, the disk can be either optically thin or optically
thick to X-rays, and we should explore both cases.

We can also calculate the accretion rate through the disk as
$\dot{M}_d \simeq 2\pi R \Sigma \alpha c_s (H/R)^2$, which yields
\begin{equation}
\dot{M}_d \sim 10^{-10} \frac{\msun}{\hbox{year}}\; \alpha n_{11}
r_4^{7/2} T_2^2\;.
\label{mdot}
\end{equation}
Note that (i) this accretion rate is extremely small; (ii) it is not a
constant with radius since the disk is not a steady state one and of
course in a real disk density $n_{11}$ will vary with
radius. Nevertheless this equation clearly shows that even if
viscosity parameter $\alpha$ is reasonably large, i.e. $\alpha\sim
0.1$, the disk is extremely dim.

\del{Pollack et al. (1994) show that the Rosseland mean opacity of
dusty accretion disks at temperature $T=100$ K is about $k_R \simeq 1$
cm$^2$/g, and increases to $\sim$ few cm$^2$/g for $100 \simlt T
\simlt 600$ K. }

\subsection{Summary of disk properties}\label{sec:dsum}

Using quiescent \sgra spectral constraints, we found that the disk
temperature should be $\td \sim 100$ K. The disk density should be
small enough to preclude it from being self-gravitating and large
enough to accommodate the mass deposition from the hot flow settling
onto the disk. The respective maximum and minimum values of the disk
density are (according to equations \ref{dn} \& \ref{nm}) are $\sim
\hbox{few} \times 10^{13}$ and $\sim \hbox{few} \times 10^{10}$
cm$^{-3}$ at radius $R = 3\times 10^4 R_g$. If wind circularization
radius is $R_c=10^4 R_g$ then these values increase to $\sim 10^{15}$
and $\sim 2 \times 10^{12}$ cm$^{-3}$, respectively. With this range
in disk density and also possibly large dynamical range between the
inner and outer radii the disk can be either optically thin or thick
in both X-rays and NIR.  If the disk is optically thick in NIR, then
most likely there must be a relatively large inner hole ($R_i\simgt
3\times 10^3 R_g$) in the disk or else the disk would have eclipsed
(Cuadra et al. 2003; Nayakshin \& Sunyaev 2003) the S2 star (Sch\"odel
et al. 2002; Ghez et al. 2003b). However most likely the disk is
optically thin in the infrared frequencies (see \S
\ref{sec:temperature}).

\section{The role of the star cluster}\label{sec:cluster}

\subsection{The rate of star-disk crossings}\label{sec:rate}

The distribution of stars in the central sub-parsec region of the
Galaxy has been a subject of intense research for many years (see
review by Genzel 2000). Existence of a stellar cusp with density
distribution $\rho_*(R) \propto R^{-1.75}$ has been predicted by
Bahcall \& Wolf (1976). Alexander (1999) found that a power-law
stellar density distribution with exponent $p$ between 1.5 to 1.75 is
statistically slightly favored over a constant density one. With
unprecedentedly high statistic, Genzel et al. (2003) determined the
stellar density distribution to follow the law
\begin{equation}
\rho_*(R) \; = \;
\begin{cases}
\rho_{*0} (R/R_b)^{-p}\,, \; R < R_b \\
\rho_{*0} (R/R_b)^{-2}\,, \; R_b < R \\
\end{cases}\;,
\label{gen03}
\end{equation}
where $R_b=10$ arcsecond, $p=1.4\pm 0.1$ and $\rho_{*0}=1.2 \times
10^{6}$ $\msun$/pc$^3$. Note that $\rho_{*0}= 69$ $\msun$/arcsec$^3$.
This power-law distribution may persist down to the tidal disruption
radius, $R_t$, which is, according to Frank \& Rees (1976),
\begin{equation}
R_t \simeq 2.0 \times 10^{13} \; {\rm cm}\; m_*^{-1/3}\; r_*\;,
\label{fr}
\end{equation} 
where $m_*$ is the mass of the star in Solar masses. 

For convenience we choose $p=3/2$ as the resulting integrals are the
simplest and the resulting accuracy is adequate for this paper. The
mass of the stars within a given radius $R$ is
\begin{equation}
M(R) \simeq 600 \left(R_b R\right)^{3/2}\;,
\label{mrs}
\end{equation}
where both $R_b$ and $R$ are in units of arcseconds. For example, at
$R=0.1$, 1 and 10 arcsecond, we have $\sim 600$, $2 \times 10^4$ and
$6 \times 10^5 \msun$, respectively.

Number of star-disk crossings per unit time and unit area is $n_*(R)
\bar{v}_{\perp}(R) \simeq (1/\sqrt{3}) n_*(R) v_K(R)$ where $n_*(R) =
\rho_*(R)/m_*$ and $\bar{v}_{\perp}(R)$ is the average of the star's
velocity perpendicular to the disk, which we accepted to be
$1/\sqrt{3}$ of the local Keplerian value. Note that stars with both
positive and negative $v_\perp$ should be counted since both produce a
flare. Taking the integral over the disk surface area from the inner
radius to the outer one, the number of star-disk crossings per unit
time is
\begin{equation}
\dot{N}(R) = (4\pi/\sqrt{3}) \sqrt{GM_{\rm BH}} n_{*0} R_b^{3/2}
\ln\frac{R_d}{R_i} =\\ 0.5 \ln\frac{R_d}{R_i}m_*^{-1}
\label{dotn}
\end{equation}
in units of day$^{-1}$.  If $R_d/R_i = 500$, then this estimate yields
$\sim 3$ star-disk crossings per day, whereas if $R_d/R_i = 10$ then
we have $\sim 1$ crossing per day. The estimate (\ref{dotn}) is very
close to the observed rate of the X-ray flares of about $1$ flare per
day (Baganoff et al. 2003b).

\subsection{Effects of the stellar cluster on the disk}\label{sec:effects}

Under the assumption that the disk is truly inactive, i.e. its
viscosity is zero, the only mechanism for a radial mass transfer in
the disk is the angular momentum loss due to the star-disk
collisions. Ostriker (1983) calculated the rate of the angular
momentum loss by the disk assuming an isotropic star cluster. We will
use his results for the Bahcall-Wolf (1976) distribution of stars
since the latter yields $n_*(R)\propto R^{-7/4}$, i.e. roughly same as
observed (Genzel et al. 2003).  For relatively small radii that we are
interested in, i.e., $R\simlt 10^4 R_g$, the angular momentum transfer
from the star to the disk is dominated by the hydrodynamical
interaction (rather than by tidal effects) and so we set $\eta = 0$
(see equation 10 in Ostriker 1983).  Ostriker (1983) defined the
``drag time'' $t_d$ as the time for the disk at radius $R$ to loose
all of its angular momentum due to the star-disk collisions. For the
Bahcall-Wolf (1976) distribution,
\begin{equation}
t_d = \frac{0.09}{n_*(R) R_*^2 v_k(R)}\;,
\label{tdo}
\end{equation}
Note that $t_d \propto R^2$, and therefore it is always the inner disk
that can be destroyed most rapidly (accreted onto the black hole) due
to the star-disk collisions.

As in \S \ref{sec:nmin}, suppose that the age of the inactive disk is
$t_a = 10^6 t_{a6}$ years. Then the disk regions with $t_d(R) \simlt
t_a$ will be emptied out by the star cluster and thus we can define
the inner disk radius $R_i$ through $t_d(R_a)=t_a$:
\begin{equation}
R_i = R_* \sqrt{\frac{\dot{N} t_a}{0.09\, 2\pi \ln{R_i/R_t}}} = 
2.2 \times 10^3 R_g r_* \sqrt{\dot{N}_3 t_6}\;,
\label{ri}
\end{equation}
where $\dot{N}_3$ is the star-disk collision rate in units of $10^3$
year$^{-1}$, and where we set $R_i=10^3 R_g$ in the slowly changing
logarithmic factor. Note that the factor in the denominator of
equation (\ref{ri}) is $\sim 4$ so that a simple interpretation of the
result is that the disk area within radius $R_i$ approximately equals
to the area of the disk covered by the star-disk collisions in $t_a$
years. It is also worth observing that the inner disk radius depends
relatively weakly on the star-disk collision rate $\dot{N}$ and $t_a$.

Finally, the inner disk may be emptied by means other than direct
stellar impacts. For example, the S2 star is as luminous as $10^5$
Solar luminosities (Ghez et al. 2003b), and it came to within $2\times
10^3 R_g$ off the black hole (Sch\"odel et al. 2002). Clearly its
separation from the disk surface would be very small at the time of
the closest approach and the disk is heated to temperatures as much as
$10^3$ K.  The disk may become ionized enough to actually start
accreting onto the black hole (at these small radii only). Therefore
we believe that our conclusion that the inactive disk is likely to
miss its inner part is quite robust. In addition disks in other LLAGN
do have inner holes with $R_i\sim 100$ to $10^3 R_g$ (Quataert et
al. 1999; Ho 2003).

\subsection{Summary: the role of the \sgra cluster}\label{sec:ssc}

We have shown (\S \ref{sec:rate}) that the rate of star-disk crossings
is few per day for the stellar cusp distribution given by Genzel et
al. (2003) This calculated event rate is rather close to what is
observed (1 flare per day). In addition, we found that the star
impacts on the inner disk are so frequent for $R\simlt 10^3 R_g$ that
the inner disk may be simply emptied out, i.e. accreted onto the black
hole. The inner disk radius was estimated to be $R_i\sim 10^3 R_g$.

\section{Schematics of a star passage through the disk}
\label{sec:single}

\subsection{Background notes}\label{sec:bnotes}

The subject of star-disk interactions in AGN has been extensively
studied in the literature. There were two main themes in these
studies. One was the influence of the stars on the disk evolution
through the additional drag imposed by the stars on the disk
(e.g. Ostriker 1983; Norman \& Silk 1983; Hagio 1987; see also \S
\ref{sec:effects}), heating (e.g. Perry \& Williams 1993) and mass
deposition due to stripping and ablation of star's envelopes
(Armitage, Zurek \& Davies 1996). Not less important were the studies
of the influence of the disk on the distribution of the stellar orbits
in the star cluster (e.g. Norman \& Silk 1983; Vilkoviskij 1983; Syer
et al. 1991; Rauch 1995; Karas \& Subr 2001; Vilkoviskij \& Czerny
2002). Zurek, Siemiginowska, \& Colgate (1994) proposed that the wake
region behind the star (when the star leaves the disk) may be the
putative ``broad line region clouds'' of AGN. Ivanov, Igumenshchev, \&
Novikov (1998) studied the passage of a black hole through the
accretion disk. At the same time relatively little attention has been
given to the emission from a single star-disk passage. The point is
that this emission is usually weak on the background of the AGN and is
unlikely to be seen, although the effects of many star-disk collisions
may be observable (Zentsova 1983; Perry \& Williams 1993).

Furthermore, in the midplane of the dense AGN-type disks, the cooling
time behind the shock is very short, and the shocked gas reaches
temperatures of ``only'' $\sim 10^3 - 10^5$ K. Thus no significant
X-ray emission was expected. However, in the case of \sgra the disk
density is probably much lower than the typical values considered in
the literature on star-disk collisions so far. In addition, even if
the disk {\em midplane} density is high enough so that the disk is
optically thick, density becomes quite small in the disk
photosphere. If the density is low, the radiative cooling time is
long, and the shocked gas is heated to temperatures as high as $10^7 -
10^9$ K and emits X-rays. This emission is too weak in the context of
normal AGN. For \sgra, however, the combination of the extreme dimness
of its quiescent X-ray emission and its proximity to us makes the
flares observable.

\subsection{Phases of the star passage through the disk}\label{sec:phases}

\begin{figure}
\centerline{\psfig{file=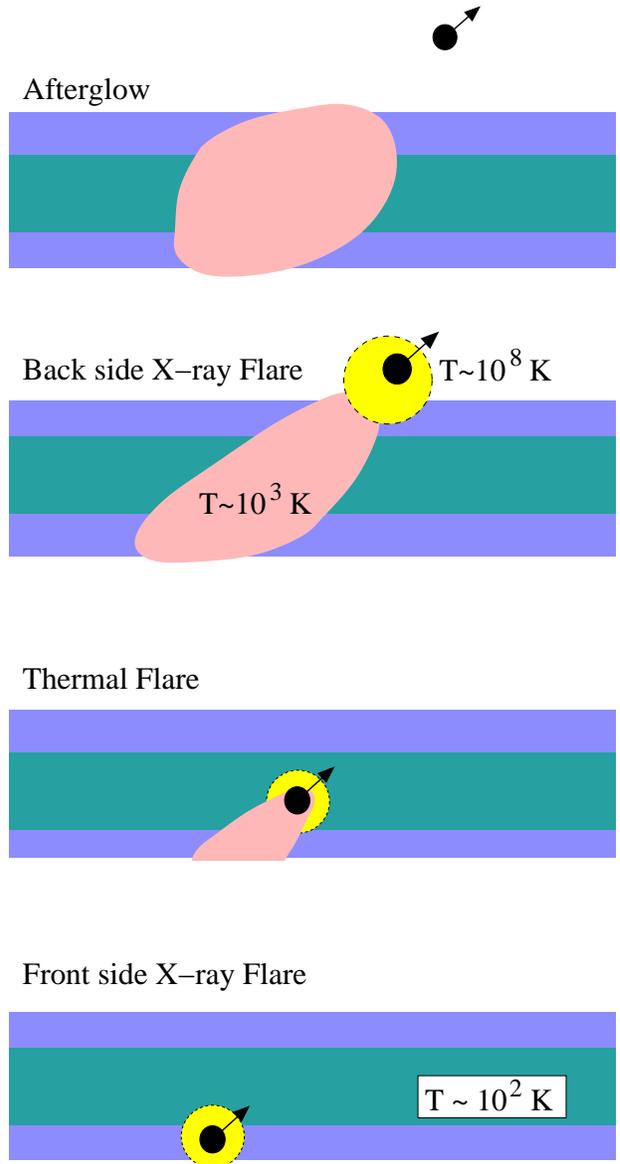,width=.45\textwidth,angle=0}}
\caption{Cartoon of a low mass star passing through the disk. Star is
shown as a filled black circle with an arrow. The disk is
schematically divided into an optically thick disk (darker) and an
optically thin -- the photosphere -- parts (lighter color). Time runs
from bottom to top. The X-rays are observable only when the star is in
the photosphere of the disk (first and third sketches from the
bottom). When the star is deep inside the disk (second sketch), the
emitted X-rays are absorbed and thermalized and emitted in the near
infra-red. In the afterglow phase (top) the hot shocked gas cooled to
low temperatures and only weak infra-red emission is present.}
\label{fig:cartoon}
\end{figure}

The star's internal density is very much larger than that of the disk
and so effects of the disk on the star during one passage are minimal
(see Syer et al. 1991). An alternative way to see this is to compare
the mass of the cold gas swept up by the star as it travels through
the disk, to the stellar mass: $\Delta M/M_* = \Sigma \pi R_*^2/M_* =
1.4 \times 10^{-11} n_{11} T_2^{1/2} r_4^{3/2} r_*^2 m_*^{-1} \ll
1$. We thus consider the star to be a rigid solid body in this
paper. We can also neglect accretion of gas onto the stellar surface
because the corresponding Bondi radius is much smaller than $R_*$
($R_{\rm B} = GM_* m_p/2k \tchar(R)= 4.5 \times 10^9 m_* r_4 \ll 7
\times 10^{10} r_*$). Tidal interactions of the star and the gas in
the wake of the star are hence neglected.

The star drives a shock into the disk. The Mach number is very large,
$v_*/c_{\rm disk}\sim \sqrt{\tchar/\td}\sim 10^3$, and thus the star
drills a very narrow (i.e. with diameter $\sim$ several $R_*$) hole in
the disk.  If the disk is optically thick, then the X-rays will be
observable only when the star, or the shock front, is in the disk
photosphere.  Figure \ref{fig:cartoon} shows the expected sequence of
events, with time increasing from the bottom to the top. The star is
shown as a black filled circle with an arrow indicating its
velocity. The optically thick denser part of the disk is shown with a
darker color, and the disk photosphere with a lighter one. For
convenience we call the side of the disk impacted by the star first
the ``front side'', and the other side as the ``back'' one.  With a
yet lighter color we show the region of the disk heated by compression
and by radiation to temperatures of order $\sim 10^3$, perhaps up to
$\sim 10^4$ K. Finally, the hottest region is the one shown with a
light colored circle around the star, with temperature of the order
the characteristic one (eq. \ref{tchar}).

Since the star's motion through the disk is highly supersonic, the
disk density remains approximately constant during the whole event
except for a small region within several stellar radii from the path
of the star. This means that the region of the disk pre-heated by
X-rays and the shock will not have enough time to expand much and form
a bulge next to the exit point of the star. The disk remains
approximately plane-parallel during the star's passage through it,
then. A good analogy to the situation is given by a bullet passing
through a low density material (e.g. a mattress).

 For the optically thick disk, the passage of the star through the
disk can be divided into three stages (cf. Figure 1):

 (1) The thermal burst stage -- the star is inside the optically thick
part of the disk.  X-rays are absorbed within the disk and re-emitted
as thermal radiation in the infra-red or optical. This is the regime
in which the star passages were extensively studied in the literature.

(2) The X-ray flare phase -- the star is in the disk photosphere. The
X-rays are directly visible to the observer in this phase. In the
simple semi-analytical treatment of this paper, we will not make
distinction between the front-side and the back-side flares, although
in reality the observed emission will depend on the 3-D geometry of
the problem through i.e. obscuration of a part of the shock by the
star itself.
    
(3) The ``afterglow'' phase -- the star is many stellar radii away
from the disk. The X-ray emitting gas cooled down to temperatures
below $\sim 10^7$ K and is not observable anymore (due to high
interstellar absorption to Sgr~A$^*$). A patch of the disk around the trek
of the star is still ``hot'' compared with the unperturbed disk, and
is cooling mostly by black-body emission coming out in the infra-red
(e.g. Syer et al. 1991).

If the disk is optically thin this division does not apply since there
is essentially only one stage.

\subsection{Division of the parameter space}\label{sec:div}

Radiation from the shocked gas is a very important facet of the given
problem since it affects both the hydrodynamics and the expected
spectra. Therefore, the parameter space for the problem naturally
breaks down into two important cases: optically thin and optically
thick (disk). We will consider these cases separately below. In
addition, there is further dependence of the results on whether the
cooling time of the shocked gas is shorter, comparable or longer than
the adiabatic expansion time (see below). Altogether we have six cases
each of which may lead to different expected flare spectra. Some of
the resulting flare characteristics, such as the duration, depends
mostly on geometry rather than on anything else.

\section{X-ray flares (luminosity)}\label{sec:xlum}

\subsection{Optically thin case}\label{sec:othin}

\subsubsection{Setup}\label{sec:setup}

Consider the case of a disk optically thin to X-rays as seen by the
observer. The photon energy window through which Chandra and XMM are
able to observe \sgra is the $E \sim 2-8$ keV range.  The opacity in
the disk photosphere is of course a very strong function of photon
energy $E$, and at small energies, i.e. $E\sim$~few keV,
photo-absorption is very significant and the opacity is much higher
than at energies $E\sim 7-8$ keV. We will consider this effect in
greater detail in \S \ref{sec:photo}, but for now we introduce an
``effective'' disk opacity, one which is a rough approximation to the
opacity weighted with the optically thin (photon energy) spectrum of
the shock. Since we expect bremsstrahlung-like hard spectra (see
below), and since photo-ionization of the gas may be important (see \S
\ref{sec:photo}), we estimate that the effective disk opacity is $b
\sim$ few times $\sigma_T$, the Thomson one.

Within this approach, the optical depth of the disk in the relevant
photon energy range along the line of sight to the observer is
\begin{equation}
\tau \simeq \frac{b}{\cos{i}} \sigma_T H \nd\;,
\label{taues}
\end{equation}
which yields the maximum midplane density for which the disk is still
optically thin:
\begin{equation}
\nd \le n_{\rm ot} = 3.86 \times 10^{11} \frac{\cos{i}}{b} T_2^{-1/2}
r_4^{-3/2} \;.
\label{ntau}
\end{equation}

Now let $R_* = r_* \rsun$ be the star's radius, $M_* = m_* \msun$ is
its mass and $\vec{v}_*$ is the star's velocity.  We define the
relative velocity of the star and the gas in the disk, $v^2_{\rm rel}
= v_K^2 + v_*^2 - 2 v_* v_K \cos\theta_r$, where $\theta_r$ is the
angle between the two velocities. In general one needs to carry out
calculations for arbitrary $v_*$ and $\theta_r$ and then integrate
over the 3D stellar velocity distribution. Here we use the simplest
approach and take the star velocity be perpendicular to the disk
surface, i.e. $\theta_r = \pi/2$. Since we also assume $v_*=v_K =
2.1\times 10^{8} r_4^{-1/2}$ cm/sec, we have $v _{\rm rel} = \sqrt{2}
v_K$.

We shall emphasize that in a more detailed calculation the role of the
exact values for $\theta_r$ and $v_*$ will be instrumental in
determining the flare characteristics. Moreover, the exact 3D
velocity of the star will also be important. For example, if the star
is moving away from the observer, then the leading edge of the bow
shock is obscured from the observer by the star. This effect is not
taken into account in our calculations. Thus we should keep in mind
that the calculations below are meant to be representative and in
reality a larger variation in the spectral and temporal
characteristics of flares is expected.

\subsubsection{Flare duration and rise time}\label{sec:xdur}

If the disk height scale is much smaller than the radius of the star,
then duration of the burst is given by $t_{\rm dur} \simeq 2 R_*/v_*$:
\begin{equation}
t_{\rm dur} \sim 2 \frac{R_*}{v_*} = 670 \; {\rm sec}\; r_*
r_4^{1/2}\;,
\label{tdyn1}
\end{equation}
In the opposite case we have $t_{\rm dur} \sim H/v_*$ 
\begin{equation}
t_{\rm dur} \sim 2\;\frac{H}{v_*} = 5 \times 10^4 \; T_2^{1/2} r_4^2
\; {\rm sec}\;.
\label{tdur}
\end{equation}
Note that $H=2 R_*$ at $r_4 = 0.11 r_*^{2/3}$, so that for radii
smaller than this one equation \ref{tdyn1} should be used while for
larger radii equation \ref{tdur} is the one to use. Since the maximum
luminosity reached in the bursts can be significantly higher than the
quiescent \sgra emission, the rise time scale should be at least
several times shorter than the burst duration. However it cannot be
much shorter than a fraction of $R_*/v_*$, i.e. a hundred seconds for
$r_* r_4^{1/2} \sim 1$.

\subsubsection{Characteristic temperature and time scales}

The maximum temperature to which the gas can be heated in the shock is
the ``characteristic'' temperature defined as
\begin{equation}
\tchar = \frac{2}{3} \frac{\mu v_{\rm rel}^2}{2 k} = 1.8 \times
10^{8} r_4^{-1}\; {\rm K},
\label{tchar}
\end{equation}
where we assumed that the mean particle mass in the shocked gas is
$\mu = m_p/2$.

In a strong shock, the gas is compressed by the factor of $4$, so that
the electron gas density in the shocked gas is $n_{\rm sh} = 4
\nd$. For the temperatures of interest the cooling function is
dominated by the bremsstrahlung radiation. Accordingly, $\Lambda(T) =
2.4\times 10^{-27} T^{1/2}$ erg cm$^{3}$ sec$^{-1}$ for a gas of
cosmic abundance (we will include atomic emission processes in the
calculation of spectra later). The cooling time of the shocked gas is
\begin{equation}
t_c = \frac{3 k T}{\Lambda(T) n_{\rm sh}} = 5.6 \times 10^3 \;{\rm sec} \;
r_4^{-1/2} n_{11}^{-1} \;,
\label{tcool}
\end{equation}
where we put $T=\tchar(R)$.  The radiative cooling time $t_c$ should
be compared with the adiabatic losses time scale, $t_{\rm ad}$, which
is the time it takes for the shocked gas to cool off via adiabatic
expansion in the disk.
\begin{equation}
t_{\rm ad} = \frac{R_*}{v_*} =\; 330 \;\; \hbox{sec}\; r_*
r_4^{1/2}\;.
\label{tad}
\end{equation}

If the gas density is very high, so that $t_c\ll t_{\rm ad}$, then the
shocked gas will not be heated to the maximum temperature
(eq. \ref{tchar}) because of the radiative cooling. We estimate that
in this case the actual temperature of the shocked gas near the star
is
\begin{equation}
T_{\rm sh} = \;\tchar\; \frac{t_c}{t_{\rm ad} + t_c}\;.
\label{tsh1}
\end{equation}

\subsubsection{X-ray luminosity}\label{sec:xlumot}

The star is essentially a piston pushing the gas ahead (and aside) of
it, and hence the rate at which the star makes work on the gas in the
disk is approximately
\begin{equation}
L_w \sim \pi R_*^2 m_p \nd v_{\rm rel}^3 \simeq 7 \times 10^{34} \;
{\rm erg\; sec}^{-1}\; n_{11} r_*^2 r_4^{-3/2}\;,
\label{work}
\end{equation}
where $n_{11}=\nd/10^{11}$ cm$^{-3}$. This is a mechanical power, and
only a fraction of this energy may be emitted in X-rays.  Defining
$\epsilon_x$ to be the efficiency with which mechanical power is
radiated away in X-rays with $E\simgt 1$ keV, we write
\begin{equation}
L_{\rm xe} = \epsilon_x L_w\;,
\label{lxe1}
\end{equation}
where the subscript ``xe'' stands for ``X-ray emitted''.  The
efficiency $\epsilon_x$ depends on the ratio of $t_c$ and $t_{\rm
ad}$. Below we use rather rough estimates of $\epsilon_x$; better
values are unlikely to be obtained in an analytical way.


If $t_c\ll t_{\rm ad}$, then shocked gas temperature (equation
\ref{tsh1}) is smaller than $\tchar$ due to strong radiative cooling.
If $T_{\rm sh}$ is lower than $\sim 10^7$ K, then most of X-ray
emission will be emitted at soft X-ray energies. Such emission would
be absorbed in the $\sim 10^{23}$ hydrogen nuclei per cm$^2$ column
depth of the cold material in the line of sight to Sgr~A$^*$. In other
words flares with $T_{\rm sh} \simlt 10^7$ K are not observable from
\sgra location. However we find that the disk midplane density that
explains the observed luminosity of flares in \sgra best is low enough
that this situation ($T_{\rm sh}\simlt 10^7$ K) is not reached. It is
then appropriate to set $\epsilon_x\simeq 1$ in the radiatively
efficient case.

In the opposite limit of a very long cooling time, the shock becomes
an inefficient radiator. The X-ray luminosity of the shock front
around the star itself is $L\sim L_w t_{\rm ad}/(t_{\rm ad} + t_c)\sim
L_w t_{\rm ad}/t_c$.  Thus in general we write
\begin{equation}
\epsilon_x = \; \frac{t_{\rm ad}}{t_{\rm ad} + t_c}\;.
\label{ex2}
\end{equation}
In particular, in the case of $t_c\gg t_{\rm ad}$, we have
\begin{equation}
L_{\rm xo} \sim 4.1 \times 10^{33} \; {\rm erg\; sec}^{-1}\; n_{11}^2
r_*^3 r_4^{-1/2}\;.
\label{lxo}
\end{equation}

It is important to note that equations \ref{ex2} and \ref{lxo} do not
include contribution of the shocked gas in the wake of the star. When
radiative cooling is initially unimportant, the shocked gas 
cools first via adiabatic losses and work against the ambient
unshocked gas. As the hot gas expands, the adiabatic cooling time
scale $t_{\rm ad}$ eventually becomes comparable to the radiative
cooling time $t_c$, at which time most of the radiation would be
emitted. If the gas temperature at that point is still larger than
$10^7$ K, then the X-ray luminosity could be as large as
$L_w$. Unfortunately we found no reliable analytical way to accurately
calculate this effect and we leave this task to future work,
remembering the fact that $\epsilon_x$ is somewhat underestimated in
the $t_c\gg t_{\rm ad}$ case.

\del{In
particular, during the adiabatic expansion of the cylindrical shell
$nr^2=$ const, where $n$ is the gas density and $r$ is the cross
sectional radius of the cylinder. For $\gamma=5/3$ gas we have
$n\propto T^{3/2}$, and therefore $T\propto r^{-4/3}$. Now, we also
have $t_{\rm ad} \propto r/\sqrt{T} \propto r^{5/3}$. For
bremsstrahlung cooling, $t_c \propto T^{1/2}/n \propto r^{4/3}$, so
the ratio of the radiative to the adiabatic time scales behaves as
$r^{-1/3}$. Furthermore, when the gas cools down to $T\sim 4\times
10^7$ K, the line cooling becomes the dominant cooling mechanism with
the cooling function behaving approximately as $\Lambda(T)\propto
T^{-0.6}$ (e.g. McKee \& Cowie 1977, Table 1). In that case $t_c$ is
nearly independent of the cylinder radius $r$ and the radiative
cooling catches up with the adiabatic one very quickly.}

\subsection{Optically thick disk case}\label{sec:othick}

We now assume that the disk optical depth to X-rays as seen by the
observer is greater than unity, and hence (cf. equation \ref{ntau})
\begin{equation}
\nd > n_{\rm ot} = 3.86 \times 10^{11} \frac{\cos{i}}{b} T_2^{-1/2}
r_4^{-3/2} \;.
\label{nthick}
\end{equation}
The main difference from the optically thin case is the fact that
X-rays emitted by the shock while the star is in the disk midplane are
absorbed and reprocessed into thermal radiation. As explained in \S
\ref{sec:bnotes}, this is the regime studied intensively in the
previous literature. However even in this case there will be X-ray
emission when the star reaches the disk photosphere. There the density
is usually low enough to preclude radiative cooling from reducing
shock temperature to below $10^7$ K. Hence the ratio of the fraction
of energy emitted in X-rays to that in the optical-UV in the optically
thick case is
\begin{equation}
\frac{\int dt \; L_{\rm x}}{\int dt \; L_{\rm ouv}}\le
\;\frac{1}{\tau_d}\; \ll 1\;,
\label{lxluv}
\end{equation}
where $\tau_d$ is the optical depth of the disk to X-rays.  Note also
that {\em two} X-ray flares are emitted per each star-disk encounter
-- one at the impact side and the other at the exit side of the disk
(see Fig. \ref{fig:cartoon}), but of course the observer will see only
the flare emitted from the side of the disk facing the observer.

The flare duration is now at most $\sim H/v_*$ because the flare from
only one side is seen. In addition, the duration will be reduced by a
further factor of a few if the photosphere of the disk is at several
hight scales above the disk (since the disk density drops very quickly
there; see eq. \ref{gauss}).

\del{Thus in general the duration of
the thermal flare, $\sim 2 H/v_*$, is few times longer than that of
the X-ray flare for optically thick disks. If the star is moving
towards the observer, the thermal flare will precede the X-ray
flare. In the opposite case the converse is true (see Fig. 2 in
Nayakshin \& Sunyaev 2003).}

\subsubsection{X-ray light curve}\label{sec:xlumthick}

The emitted X-ray luminosity can be calculated exactly as in the
optically thin case (\S \ref{sec:xlumot}) but we now take into account
(i) attenuation of X-rays in the disk; and (ii) the fact that
unshocked gas density in the disk photosphere, $n_{\rm p}$, is lower
than that in the disk midplane, $\nd$. Our treatment of the first
complication is to simply introduce exponential attenuation along the
line of sight. This approach underestimates the X-ray luminosity for
very small $\cos{i}$ for which radiation scattered once in the
photosphere becomes more important than the unscattered one.  We defer
a more careful treatment of the scattered radiation to a future paper
however, to keep our calculations here as analytical and transparent
as possible.

One should also keep in mind that we use the ``effective'' cross
section for the total one (photo-absorption plus scattering), which we
parameterize as $b \sigma_T$ (see \S \ref{sec:setup}). Now let
$\tau_e$ be the effective total depth of the disk from the star's
position $z$ to the observer. The {\em observed} X-ray luminosity is
then
\begin{equation}
L_{\rm xo} \sim L_{\rm xe}\; \exp\left[-\tau_e\right]\;,
\label{lxbol}
\end{equation}
where $L_{\rm xe}$ is given by equation \ref{lxe1}.  The gas density
in the photosphere is approximately given by the gaussian density
profile:
\begin{equation}
n(z) = \nd \exp\left[-\left(\frac{z}{H}\right)^2\right]\;.
\label{gauss}
\end{equation}

The observed X-ray luminosity is now a function of time since $z =
v_{*z} t$ if we define $t=0$ as the time at which the star crosses the
disk midplane. Consider a front side flare -- when the star is moving
into the disk (i.e., see the sketch on the very bottom of
Fig. \ref{fig:cartoon}; the observer is below the disk, at $z=-\infty$
in this case). When the star first impacts the disk photosphere, the
X-ray luminosity rapidly increases because the star moves into denser
and denser environment. During this optically thin stage $L_{\rm
xo}\simeq L_{\rm xe} \propto \epsilon_x n(z)$ (equation
\ref{lxe1}). Later, when the surface $\tau_e=1$ is reached, the
attenuation of X-rays by the disk matter becomes important and the
luminosity dependence on the density becomes $L_{\rm xo}\propto
\epsilon_x n(z) \exp(-\tau_e)$. Furthermore, for the gaussian density
profile
\begin{equation}
\tau_e(z) \;\simeq\; \frac{n(z)}{\nd}\; \tau
\label{tauz}
\end{equation}
where $\tau$ is the optical depth to the disk midplane (equation
\ref{taues}) and $\nd$ is the disk midplane density. It then follows
that
\begin{equation}
L_{\rm xo}(t)\;\propto \epsilon_x \;n(z)\;\; \exp\left[-\tau\;
\frac{n(z)}{\nd}\right]\;.
\label{lxen}
\end{equation}
Thus the event is indeed a flare: the luminosity first increases and
then decreases quite rapidly. Note however that we implicitly assumed
here that cooling time $t_c$ is shorter than duration of X-ray flare,
$t_{\rm dur}$.

\subsection{Maximum flare X-ray luminosity}\label{sec:xmax}

The maximum observed X-ray luminosity, $L_{\rm max}$, for an optically
thick disk, will be approximately that corresponding to
$\tau_e=1$. The gas density at which $\tau_e=1$ is simply $n(z)=
n_{\rm ot}$ as given by equation (\ref{ntau}). We now substitute this
value for the gas density into equation \ref{lxe1} and multiply the
result by $\exp[-1]$ to get
\begin{equation}
L_{\rm max} = 10^{35} \; \epsilon_x(n_{\rm ot})\; \frac{\cos{i}}{b}\;
T_2^{-1/2} r_*^2 r_4^{-3}\;,
\label{lmax}
\end{equation}
where $b\sim$ few is the ratio of the effective total cross section in
the disk atmosphere to the Thomson cross section.  Note that in the
case of an optically thin disk the midplane density is lower than
$n_{\rm ot}$ and hence the emitted X-ray luminosity is
lower. Therefore equation \ref{lmax} gives the maximum X-ray
luminosity of a flare in both optically thin and thick cases.

We shall emphasize the fact that the maximum X-ray luminosity just
derived is a strong function of the disk inclination angle $i$. In
particular in the radiatively inefficient shock case,
$\epsilon_x\propto \cos{i}$ and thus $L_{\rm max}\propto
\cos^2{i}$. This dependence shows that if X-ray flares are indeed
associated with the putative inactive disk, then the inclination angle
could not be too large, i.e. the disk could not be face on or else the
X-ray luminosity of the flares would be too small.

\section{X-ray spectrum}\label{sec:semit}

\subsection{Optically thin spectrum}\label{sec:sot}

In general the X-ray spectrum of a flare is complicated. The emitting
region is a multi-temperature gas and only numerical calculations are
likely to yield accurate spectral predictions. However single
temperature optically thin spectra will provide us with certain
guidance. In addition, these spectra should be of some relevance for
the optically thick disk and the case of $t_c \sim t_{\rm ad}$. Under
these conditions we expect $T_{\rm sh}\sim \tchar$ and a major
fraction of the mechanical power of the shock should be ``quickly''
radiated in X-rays, so that the wake of the star does not make a large
contribution to the emitted spectrum.

The photo-ionization parameter for the gas in the shock is $\xi \simeq
L_x/n_{\rm sh} R^2$ and is usually smaller than a hundred, i.e.,
photo-ionization is not important for the shocked gas, and the plasma
is essentially in the coronal limit. The ionization equilibrium
assumption is appropriate since the recombination time scale for the
completely ionized Fe is $1/(\alpha_{\rm rec} n_{\rm sh}) \sim 3$ sec
$n_{11}^{-1} T_7^{1/2}$.  Further, the Thomson optical depth of the
shocked gas is $\tau_T \sim \sigma_T n_{\rm sh} R_* = 0.02 r_* n_{11}
\ll 1$, so that the emission region is optically thin. The Compton
$y$-parameter, $y\equiv 4 (kT_{\rm sh}/m_ec^2) \tau_T \ll 1$ in most
realistic cases and thus Comptonization is not important for the
continuum.

\begin{figure}[t]
\centerline{\psfig{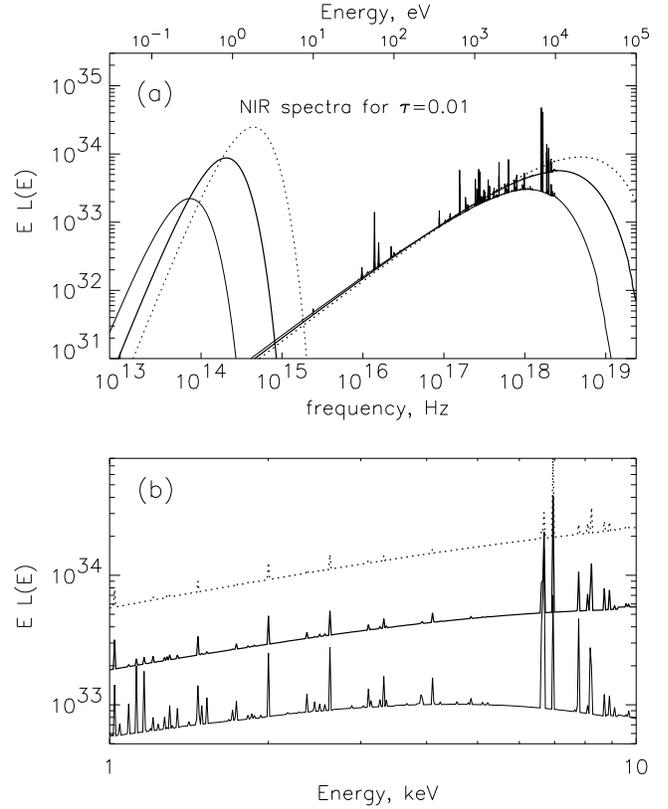}}
\caption{Optically thin X-ray spectra for three different values of
disk radius: $r_4 = 0.5, 1\; {\rm and}\; 2.5$ for the dotted, solid
thick and thin curves, respectively. The gas density is $\nd =
10^{11}$ cm$^{-3}$, inclination angle $i=70^\circ$, and the star's
radius is $R_* = 2 \rsun$ for all the three cases. (a) -- Spectra in a
broad frequency range. The near infra-red curves will be explained
later in \S \ref{sec:nir}. (b) -- the X-ray part of the spectrum. The
upper curve is shifted up by 3 whereas the lower one shifted down by
the same factor for clarity. Note that, except for the brightest of
the curves, the ``hot'' \fe line is clearly present.}
\label{fig:allsp}
\end{figure}

We computed optically thin X-ray spectra using the code XSTAR (e.g.,
Kallman \& Krolik 1986), assuming solar abundance of elements and that
the emitting gas temperature is equal to $\tchar(R)$. The results for
three different radii, $r_4 = 0.5$, 1 \& 2.5, are plotted in Figure
\ref{fig:allsp} with the dotted, thick and thin solid curves,
respectively. The near infrared part of the spectra will be discussed
later in \S \ref{sec:nir}. In Figure \ref{fig:allsp}, the top panel
shows the spectra in a broad frequency range and the bottom panel
concentrates on the $1-10$ keV part of the spectrum.

The characteristic temperatures are $\tchar = 3.6, 1.8\; {\rm and}\;
0.73 \times 10^8$ K for $r_4 = 0.5, 1\; {\rm and}\; 2.5$,
respectively. For such high temperatures, the X-ray spectrum is mostly
bremsstrahlung emission with the great exception of the \fe lines at
$\simeq 6.7$ and $6.9$ keV. The keV-part of the spectrum shows that
the line is dominated by the He-like (i.e., the $6.7$ keV) component
for the smallest $T$ considered, whereas the 6.9 keV component
dominates for the largest value of $T$.

\begin{figure}
\vskip0.5cm
\centerline{\psfig{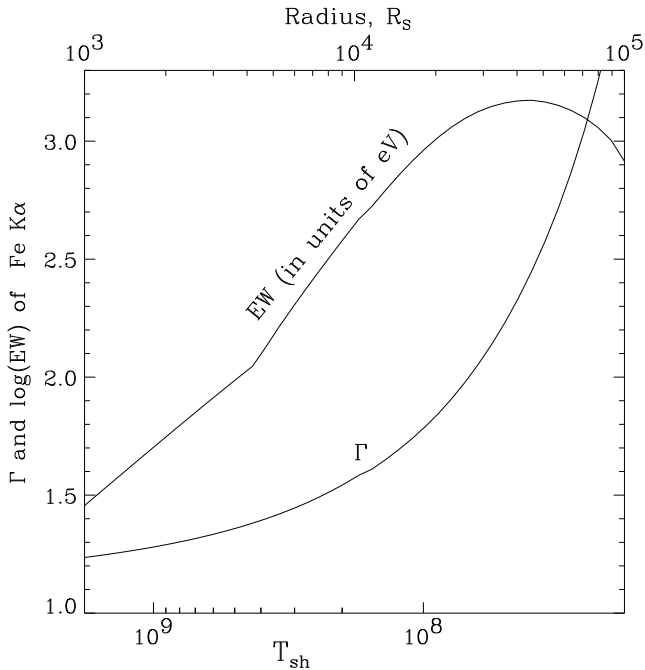}}
\caption{(a) $\log$(EW) of \fe line and the X-ray spectral index
$\Gamma$ versus temperature $T_{\rm sh}$ of the emitting gas for
optically thin X-ray spectra for Solar abundance of elements.  If
$T_{\rm sh} = \tchar(R)$, then $\Gamma$ and EW may be plotted versus
radius $R$, shown as the upper $x$-axis.  Note that to yield \fe EW
and spectral index $\Gamma$ compatible with the observations of the
two bright flares (Baganoff et al. 2001 \& Goldwurm et al. 2003a),
$T_{\rm sh} \simgt 2\times 10^8$ K or $R\simlt 10^4 R_g$.}
\label{fig:gamma}
\end{figure}

Clearly, the integrated equivalent width (EW) of \fe line decreases as
$\tchar$ increases ($R$ decreases).  Another obvious trend seen in
Figure (\ref{fig:allsp}) is that with increase in $\tchar$ the thermal
rollover moves to higher frequencies and the spectrum begins to look
more and more power-law like in the $2-8$ keV band. To quantify these
trends and to compare the observed flare spectra with those predicted
by our model, we defined the photon index $\Gamma$ by drawing a
power-law through two line-free photon energies of $E\simeq 2.2$ keV
and $7.25$ keV, and also integrated over all the components of the \fe
line to get its total EW. Figure \ref{fig:gamma} displays the
resulting dependence of index $\Gamma$ and the EW of \fe line on the
temperature of the shocked gas.

So far only two -- the brightest -- of the observed X-ray flares had
enough statistics to conclude that spectra during flares seem to be
hard power-laws in the $2 \simlt E \simlt 8$ keV energy range, with
the photon spectral index $\Gamma\simeq 1\pm 0.7$ (Baganoff et
al. 2001, Goldwurm et al. 2003a), and they seem to show little of an
\fe line. Goldwurm et al. (2003b) notes that a \fe line-like feature
at $\sim 6.4$ keV in their spectra is not statistically significant;
but the upper limit on the EW of the line is as large as 1.8 keV. From
Figures (\ref{fig:allsp} \& \ref{fig:gamma}) it is clear that to have
such hard spectra, the bright flares should have $T_{\rm sh} \simgt 2
\times 10^8$ K.  The constraint on the EW of the \fe line also
requires similarly high values of $T_{\rm sh}$.

Finally, let us recall that in the case $t_c\gg t_{\rm ad}$, we expect
that there will be a range of temperatures that contribute to the
overall spectrum, from $T=\tchar$ in front of the star to much lower
temperatures in the wake of the star. The main effect is to make the
spectrum softer. Hence the minimum value of $\Gamma\simeq 1.3$ from
Figure \ref{fig:gamma} should also be the minimum value for the
spectral index in the more general case.

\subsection{Absorption along the line of sight in the photosphere}\label{sec:photo}

The emitted spectrum is attenuated in the atmosphere {\em
differentially} for different photon energies $E$ because photon
opacity is a function of $E$. The observed spectrum could be severely
cut (compared to that shown in Fig. \ref{fig:allsp}) at low energies
$E\simlt 2$ keV because the opacity there is very large. Observations
of at least the brightest flares show no enhancement in the column
depth of the {\em cold neutral} absorber in the line of sight
(Baganoff et al. 2001, Goldwurm et al. 2003a).  For optically thick
case, the only way to reconcile the model with the observations is if
the photosphere around the star is strongly ionized and hence the
absorption opacity is much smaller than that due to the cold material
in the line of sight.

\begin{figure}
\centerline{\psfig{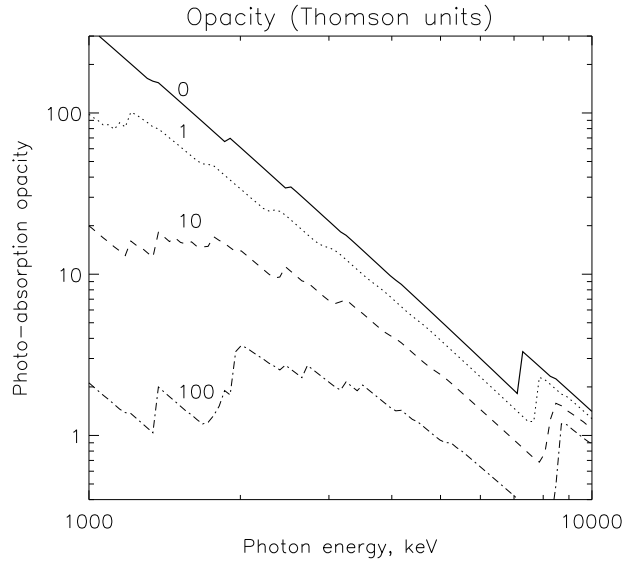}}
\caption{Photo-absorption cross section (in units of $\sigma_T$) of
photo-ionized matter as a function of photon energy. Values of
photo-ionization parameter $\xi$ are indicated next to the respective
curves. Note a very fast decrease in the opacity with increase in the
ionization parameter for $\xi \simgt 1$. This occurs because of
photo-ionization of the elements providing the opacity.}
\label{fig:opacity}
\end{figure}

Baganoff et al. (2001) found that the column depth of the cold
material during both quiescent and flare emission is $N_{\rm cold}
\sim 5 \times 10^{22}$ hydrogen nuclei per cm$^2$. The column density
through which the X-rays pass during the maximum X-ray observed
emission is around $N_{\rm ion} \sim 1/b\sigma_t \sim 5 \times
10^{23}$. We then need the absorption opacity of the ionized
photosphere to satisfy
\begin{equation}
\frac{\sigma_{\rm ion}(E)}{\sigma_{\rm cold}(E)} \ll \frac{N_{\rm
cold}}{N_{\rm ion}} \sim 0.1 \;,
\label{dsigma}
\end{equation}
where $\sigma_{\rm cold}$ is the absorption cross section of the cold
gas which is much larger than $\sigma_T$, the Thomson cross section.
For energy $E$ in equation \ref{dsigma} it is sensible to take $E\sim
2$ keV because the cold gas opacity is largest at the smallest
energies in the keV band, and so the measurement of $N_{\rm cold}$ is
really dominated by the lowest energy bin (see Figure 3 in Baganoff et
al. 2001).

The ionization state of the disk atmosphere illuminated by X-rays from
the shock is controlled by photo-ionization rather than by collisional
ionization since the gas temperature is low, i.e. $T\sim T_{ir}\sim
10^3 - 10^4$ Kelvin. Using XSTAR once again, we calculated the
ionization state of plasma kept at $T=10^4$ K and illuminated by
X-rays with spectrum similar to those shown in Figure
(\ref{fig:allsp}). The resulting absorption cross section as a
function of photon energy is plotted in Figure (\ref{fig:opacity}) for
different values of photo-ionization parameter which is defined as
$\xi\equiv 4\pi F_x/n_0$ where $F_x$ is the integrated X-ray flux. The
solid curve labelled ``0'' is calculated for a very small $\xi$ and
yields the ``cold'' absorption cross section $\sigma_{\rm cold}(E)$.
The Figure shows that the condition specified by equation
(\ref{dsigma}) is fulfilled when ionization parameter exceeds about
10. 

We estimate the average ionization parameter in the photosphere around
the star as 
\begin{equation}
\xi \sim \frac{L_{\rm xe}}{n_0 (H/2)^2}\;,
\label{xim}
\end{equation}
where $L_{\rm xe}$ is the emitted X-ray luminosity at the maximum
light, which is given by $L_{\rm max}\;{\rm e}^1$ (see \S
\ref{sec:xlum}), and as representative distance from the star $H/2$ is
chosen. For example, $L_{\rm max}$ for the case of an optically thick
disk with adiabatic losses dominating the radiative ones is
\begin{equation}
\xi \sim 0.042 \frac{\cos{i}}{\sigma_e} T_2^{-3/2} r_*^3
r_4^{-5} \;.
\label{xic}
\end{equation}
Clearly strong enough ionization of $\xi\simgt 10$ requires the two
bright detected X-ray flares to happen at small values of $r_4$ and/or
be produced by large stars with $r_*\gg 1$. Taking a concrete example
of a 5 Solar radii star, we find that $\xi$ exceeds 10 for radii $r_4
\simlt 0.6$.  Note that the photo-ionization parameter $\xi$ at the
time of maximum light in X-rays is a function of the disk inclination
angle, so once again we see the dependence of the results on the disk
orientation to us (cf. \S \ref{sec:xlumthick} and \ref{sec:setup}).

We shall emphasize that the constraint $\xi\simgt 10$ is important
only if the disk is somewhat optically thick. The disk may be
optically thin to X-rays which we find quite likely for the innermost
disk region where the bright flares should be taking place.

\section{Flares in the near infrared}\label{sec:nir}

Here, as everywhere else in this paper, we concentrate on the effects
of relatively low mass stars on the disk. For high mass, high
luminosity stars, the main effect through which the disk is affected
is the heating of the disk by the enormous optical and UV radiation
output of these stars (Cuadra et al. 2003). In addition winds from
such stars may increase the effective radius of the star: due to the
winds a larger area of the disk may be shock heated.  These stars
should be expected to produce NIR and X-ray flares much stronger than
those we consider in this paper. Of course such powerful flares are
much less frequent than those from low mass stars, and we plan to
consider in our future work.

\subsection{Optically thick case}\label{sec:tnir}

First consider the simplest case of a disk optically thick in the near
infrared. X-rays emitted by the shock while inside the optically thick
disk are easily absorbed by the cold un-shocked gas and reprocessed
into thermal radiation. Therefore we assume that all the work done by
the star on the cold gas is emitted as a blackbody radiation in this
case and that the luminosity is equal to $\simeq L_w$ (equation
\ref{work}). The radiation runs ahead of the shock wave, heating and
ionizing the gas in the disk.  The region affected by the radiation is
roughly the cube with size $\sim 2 H$. The temperature of the region
can be found by assuming that the cooling time within the disk,
$t_{cd}$, is shorter than $H/v_*$.  This implies that $ L_w \sim
\sigma_B T_{ir}^4 \times 2 (2 H)^2$ and
\begin{equation}
T_{ir} = 1.78 \times 10^3 \;{\rm K}\; \left[\frac{n_{11}r_*^2}{T_2
}\right]^{1/4} r_4^{-9/8}\;
\label{tir}
\end{equation}
(index $ir$ stands for ``infra-red'').  If $t_{cd} > t_{\rm pre}$,
then the thermal luminosity is lower than $L_w$.

As remarked in \S \ref{sec:single}, the pre-heated gas cannot expand
much during the star's passage through the disk. The observed flare
luminosity in near infra-red is then reduced by the disk projection
effect: $L_{\rm obs} = L_w \cos{i}$. The near infra-red K-band seems
to yield the tightest constraints on the theoretical models (see,
e.g., Hornstein et al. 2002). The corresponding reference wavelength
is $2.2 \mu$m and the frequency is $\nu_0 = 1.36 \times 10^{14}$
Hz. The predicted observed luminosity is then $\nu L_{\nu} = 4 \pi
\times 4 H^2
\cos{i}\; \nu B_{\nu}(T_{ir})$:
\begin{equation}
\nu L_{\nu} = 7.8 \times 10^{36} \cos{i}\; T_2\, r_4^3\;
\frac{\bar{\nu}^4}{e^x - 1}\;,
\label{lnir}
\end{equation}
where $\bar{\nu}\equiv \nu/\nu_0$ and $x\equiv h\nu/kT_{ir}$, and
where we assumed that the distance to the Galactic Center is $D=8.0$
kpc. Note that the factor of $4 \pi$ above arises due to the usual
assumption of the observed at Earth that the emitter (the disk) emits
isotropically in the full $4\pi$ steradians. Also note that the
dependence on the disk density is ``hidden'' in the temperature
$T_{ir}$ of the emission and that $x = \bar{\nu}\; (h\nu_0/kT_{ir}) =
\bar{\nu}\; (6.5 \times 10^3 {\rm K}/T_{ir})$.

The luminosity given by equation \ref{lnir} is quite large. The
current best limit on the quiescent emission of \sgra at 2.2 $\mu$m is
about $2$ mJy (deredenned; Hornstein et al. 2002). In terms of $\nu
L_{\nu}$, the quiescent limit of 2 mJy corresponds to about $2 \times
10^{34}$ erg/sec. Hornstein et al. (2002) note that the limits on
random 3-hour flares is about a factor of 10 higher, i.e. $\simeq
2\times 10^{35}$ erg/sec. Hence an optically thick (in NIR) disk in
\sgra would produce NIR flares strong enough to be detected by now. 

\subsection{Optically thin disk}\label{sec::onir}

Previously we pointed out that no optically thick disk can exist in
\sgra already based on its quiescent spectrum -- or the disk would
reprocess too large a fraction of the visible and UV stellar flux into
the infrared band and would violate the observed upper limits (see \S
\ref{sec:temperature}). These constraints limit the infrared optical
depth of the disk to $\tau_{\rm ir}\simlt 0.01$. If the opacity of the
disk grains were grey, the grain temperature would equal to the
effective one (that is $T_{ir}$). In this approximation the luminosity
of the flare in the NIR is simply reduced by the factor $\tau_{\rm
ir}$:
\begin{equation}
\nu L_{\nu} = 7.8 \times 10^{34} \frac{\tau_{\rm ir}}{0.01}\;\cos{i}\;
T_2\, r_4^3\; \frac{\bar{\nu}^4}{e^x - 1}\;.
\label{lnir2}
\end{equation}
Figure \ref{fig:allsp}, panel (a), shows examples of such spectra for
a star with radius $R=2 \rsun$. Note that the NIR flares are weak
enough to be missed. Also note that the value of the dust optical
depth, assumed to be equal to 0.01 for all frequencies, is likely to
be smaller than that of the disk in quiescence. The stellar passage
through the disk may in fact destroy the dust (see below) in which
case the NIR luminosity will be reduced even further.

It is tempting at this point to make a connection between $\Sigma$,
the disk surface density given by equation \ref{sigma}, and the
optical depth in the NIR using the standard interstellar dust
opacity. For example, at $2.2 \mu$m, the dust extinction is about $5
\times 10^{-22}$ cm$^2$ per Hydrogen atom (see Fig. 2 in Voshchinnikov
2003). Thus with $\nd\sim 10^{11}$ cm$^{-3}$ the disk would be
optically thick in the $2.2 \mu$m for radii greater than about $10^3
R_g$. However the dust opacity in the disk is likely to be reduced
compared with the interstellar values but it is hard to calculate this
process model-independently.  One uncertain quantity is the minimum
size of the dust grains which is influenced by dust growth. For
example, following Spitzer (1978), we estimate that the grain radius
$a$ grows at the rate $da/dt \simeq 10^{-3} \zeta_a n_{11}^{-1}$
cm/year, where $\zeta_a\simlt 1$ is the average fraction of atoms that
stick to the grain in the course of collisions with the grain. This
rate is very large so there may be no small grains in the disk which
would argue for an optically thin disk. On the other hand dust grains
will be destroyed or broken up into smaller grains by dust-dust
collisions.

Another uncertainty is the ratio of the mass of the matter in the
grains to that in the gas. There could be no dust in the disk
initially when the active phase turned off because the disk was too
hot; the dust could then be created in the disk and could also be
brought in by the stellar winds; it could also be destroyed because
of stellar UV radiation, star passages. For an example we briefly
consider the last effect mentioned above. The temperature of the gas
in the cylinder with radius $H$ around the star's trek in the disk is
given by equation \ref{tir}. This temperature is high enough to
destroy all or most of the dust species. Therefore we may define time
``dust destruction'' time scale, $t_{\rm dd}$, as time needed to cover
all of the disk area within radius $R$ by star-disk collisions (each
collision ``covers'' area $\sim \pi H^2$ of the disk). Using the rate
of the star-disk passages calculated in \S \ref{sec:rate}, we obtain
\begin{equation}
t_{\rm dd} = \frac{\pi R^2}{\dot{N}(R) \pi H^2} \simeq 2 \times 10^4
\;\hbox{years}\; T_2^{-1} r_4^{-1}\;.
\label{tdd}
\end{equation}
This time is very short compared with the viscous time scales for the
cold disk ($\sim 10^7 - 10^8$ years).

\section{Typical ``weak'' and typical ``strong'' flares}\label{sec:typ}

\begin{figure}
\centerline{\psfig{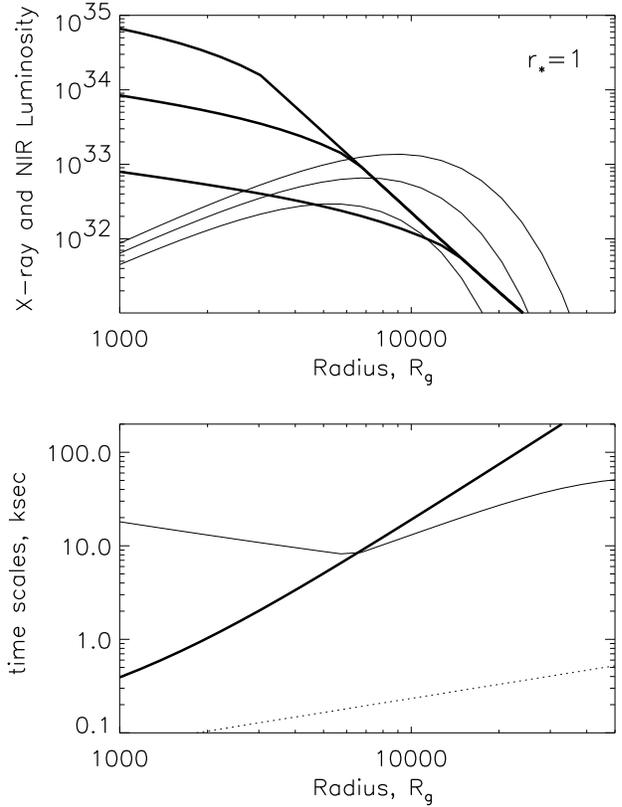}}
\caption{Flare properties for $R_* = \rsun$, disk inclination angle of
$i=75^\circ$ and $b=2$ (see eq. \ref{lmax}). (a) -- Maximum X-ray
(thick) and NIR (thin) luminosity during flares as a function of
radius for three different values of disk midplane densities, $\nd =
3\times 10^{10}$, $10^{11}$ and $3\times 10^{11}$ hydrogen nuclei per
cm$^3$. For NIR luminosities, we assume that the dust optical depth is
0.01, as in equation \ref{lnir2} (b) -- Burst duration (thick solid),
adiabatic expansion time (dotted) and the cooling time (thin solid).
Only the latter depends on the disk density, which was chosen to be
$\nd = 10^{11}$ for the thin solid curve.}
\label{fig:rad1}
\end{figure}

Given the theoretical picture of a star-disk collision developed in
this paper, we now consider two important cases. The first one is the
passage of a ``low''-mass star with $R_*=\rsun$ through the disk for a
range of midplane disk densities $\nd$, and for different radii $R$.
The second case is that of a ``large'' star, with $R=5 \rsun$, passing
through the same disk.

The low-mass star case is analyzed in Figure \ref{fig:rad1}. The
midplane disk density $\nd$ is chosen to be $\nd = 3\times 10^{10}$,
$10^{11}$ and $3\times 10^{11}$ hydrogen nuclei per cm$^3$.  The X-ray
luminosities are shown with thick curves while the near infrared
2.2$\mu$ luminosities are plotted with thin curves in panel
(a). Larger values of these luminosities correspond to larger values
of the midplane density. Panel (b) of the Figure shows three time
scales -- the duration (eq. \ref{tdur}), cooling (eq. \ref{tcool}) and
adiabatic expansion time (eq. \ref{tad}). Out of these only the
cooling time depends on the disk density, which was chosen to be $\nd
= 10^{11}$ cm$^{-3}$ for the respective curve in panel (b). We do not
show disk radii smaller than $10^3 R_g$ because it is rather unlikely
that the disk will extend there (see \S \ref{sec:effects}).

\begin{figure}
\centerline{\psfig{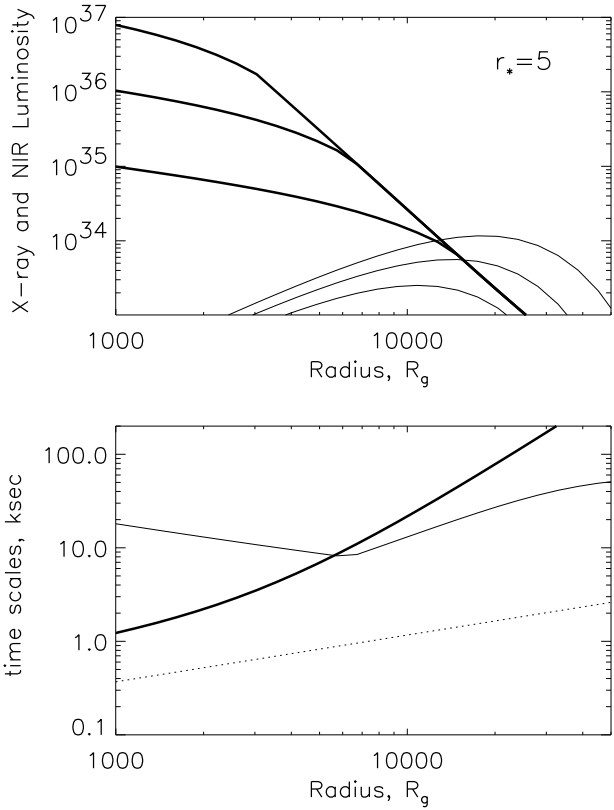}}
\caption{Same as Figure \ref{fig:rad1} but for a high mass star with
$R= 5 \rsun$. Note the change in the luminosity scale; X-ray flares
are roughly 100 times brighter than they were in Figure
\ref{fig:rad1}. The NIR emission increased by a much smaller factor
mostly because it is in the Rayleigh-Jeans regime.}
\label{fig:rad5}
\end{figure}

The X-ray luminosity curves have a common asymptotic at large radii,
which is simply the optically thick case given by equation
(\ref{lmax}). It is very important to note that in this optically
thick limit the X-ray luminosity is an extremely strong function of
radius: $L_{\rm xo}\propto R^{-3}$. Given the fact that quiescent
X-ray luminosity of Sgr~A$^*$ is about $L_q \sim 10^{33}$ erg/sec, it
is clear that flares from solar-sized stars will not be observable for
$R\simgt 10^4 R_g$. A comparison with the case of a large star, Figure
\ref{fig:rad5}, shows that flares from these are observable up to
radii of $\sim \hbox{few}\times 10^4 R_g$, i.e. just a factor of few
higher. Moreover, this conclusion is {\em independent} of the disk
midplane density: the X-ray luminosity of stars passing through the
disk at a given radii cannot exceed $\lmax$. Thus we conclude that as
far as X-ray flares are concerned, only the ``inner'' disk region with
$R < \hbox{few}\times 10^4 R_g$ matters.

At smaller disk radii, the disk becomes optically thin to X-rays along
the line of sight to the observer and at this point the X-ray
luminosity curves change their slope. The X-ray luminosity dependence
on the radius is much weaker in the optically thin regime (see
eq. \ref{lxo}). Clearly ``large'' stars produce higher luminosities
than do the ``small'' ones, in both optically thin and thick
regimes. This is easily understood from the fact that the total work
done by the star on the disk is proportional to the area of the star
(equation \ref{work}), plus radiative cooling is actually more
important for larger stars (equation \ref{lxo}). There is of course
fewer high mass stars than there is low mass ones in \sgra star
cluster (Genzel et al. 2003), and this means that the prediction of
the model is that high luminosity flares should be a minority.

Spectra emitted by flares from passages of low mass stars are likely
to be relatively soft, i.e. with $\Gamma \simgt 2$. We reach this
conclusion based on the fact that radiative cooling time for such
flares is always much longer than the adiabatic expansion time (see
thin solid versus dotted curves in Fig. \ref{fig:rad1}), and hence the
gas will expand and cool significantly before a large fraction of the
shock mechanical power is radiated away. As a result one should also
expect a strong ``hot'' \fe line to be emitted by such flares.

On the other hand, for high mass stars the adiabatic expansion time is
longer and hence the ``average'' temperature of X-ray emitting gas
should be larger than that from the smaller stars. This is especially
true for star-disk collisions at radii $\simlt \hbox{few} \times 10^3
R_g$ because then the characteristic temperature is very high,
i.e. $\tchar\sim 10^9$ K, and even if the emitting gas temperature is
significantly smaller than this it may still be high enough that the
2-8 keV part of the spectrum is a pure power-law with $\Gamma\sim 1.5
- 2$. Correspondingly \fe line emission should be much weaker from
such flares. Since these are also the brightest flares we conclude
that, statistically speaking, bright flares should have harder spectra
than the less luminous ones, and they should show less of the \fe line
emission.

Duration of typical flares is in the range of one to tens of kilo
seconds. We should stress that this result is quite model
independent. Flares with longer durations are not detectable because
they are too weak; flares shorter that about a thousand seconds may
not exist because we found that the disk cannot extend much into the
innermost region (see \S \ref{sec:rate}). Even if it did, the cooling
time there becomes very long and the X-ray emission would then
continue for times longer than $t_{\rm dur} \sim H/v_*$.

The near infrared (2.2$\mu$ here) emission is quite weak in all the
models, at or below the $\sim 2 \times 10^{35}$ erg/sec upper limit
set by Hornstein et al. (2002). This is partially because the
black-body emission is a rather narrow feature and in most cases the
2.2$\mu$ is either on the Wien exponential tail or on the
Rayleigh-Jeans part of the curve. We should also note that the NIR
luminosities are calculated under the optically thick assumption which
may be violated for the smallest values of the disk surface density
$\Sigma$. However most likely optically thin spectra would peak in the
optical band again producing little NIR flux to be detected (see \S
\ref{sec:nir}). Next ``complication'' in the way of detecting NIR
flares from the GC in the context of our model is the fact that they
happen not exactly at the \sgra position. The angular separation
correspond to $R = 10^4 R_g$ is about 0.1\arcsec. With current
telescopes a star's position can be measured with a better accuracy,
e.g. to within about 0.01\arcsec (Sch\"odel et al. 2002), and hence if
a flare indeed happened $\sim 0.1$ arcsec away from \sgra position it
would not be even counted as the one coming from the black hole and
probably be discounted as noise.

The cyclotron frequency in the hot blob is about 1 GHz, assuming
equipartition magnetic field. However, its intensity is many orders of
magnitude larger than the Rayleigh-Jeans intensity at this frequency,
so this emission is completely self-absorbed and therefore impossible
to observe. It is unlikely that non-thermal mechanisms could
significantly alter this conclusion. Therefore our model predicts no
radio-flares correlated with X-ray flares (but star-disk interactions
may have indirect implications for the jet and hence some radio
variability may in fact be due to stars; see also Nayakshin \& Sunyaev
2003).

\section{Comparison with observations of X-ray flares}\label{sec:comp}

Observational facts for X-ray flares from \sgra may be summarized as
following (Baganoff et al. 2001, 2003b; Goldwurm et al. 2003a). The
maximum X-ray luminosity detected so far is $L_x \sim 10^{35}$
erg/sec. Such strong flares are relatively rare; i.e. there is
probably $\sim 5-10$ times more of the weaker $L_x\simlt 10^{34}$
erg/sec flares. Counting all detectable flares, there is about one
flare per day on average. There seems to be no NIR or radio
counterparts to the flares (Baganoff et al. 2003b; Hornstein et
al. 2002). In particular the radio source has never varied by more
than a factor of 2 during many years of observations (Zhao, Bower, \&
Goss 2001). The NIR limits on random $\sim 3$ hour flares are about
$\sim 2 \times 10^{35}$ erg/sec, much too tight for many existing
models of \sgra (Hornstein et al. 2002). The spectra appear to be
harder than those during quiescence and show little if any \fe line
emission (both Baganoff et al. 2001 and Goldwurm et al. 2003a,b X-ray
spectra contain weak emission-line like features at energies not too
far from the expected \fe line but detailed modeling is difficult due
to poor statistic. Goldwurm et al. 2003b put an upper limit of 1.8 keV
on a 6.4 keV \fe line feature in their spectrum of an X-ray flare). In
addition there are indications that some flares may be much softer and
are complex around the \fe line energy (P. Predehl, private
communication). To this list of \sgra flare properties we should also
add the fact that if \sgra is indeed a Low Luminosity AGN, then it is
the only one that shows such large magnitude flares.

The following flare characteristics appear to have a reasonable
explanation in our model (we place items in order from very model
independent to less so):

\begin{itemize}

\item (i) Frequency of flares, i.e. number of flares per day, which in
our model is around few per day (see \S \ref{sec:rate} and in
particular equation \ref{dotn}).

\item (ii) Duration of X-ray flares is in the range of $\sim$ one to
tens of kilo seconds.

\item (iii) X-ray luminosities of observed flares in the peak of their
light-curves are consistent with the maximum allowed luminosities in
our model (equation \ref{lmax} with $\epsilon_x$ given by equation
\ref{ex2}). Namely, the observed luminosities are best explained by an
optically thin disk with midplane density of order $10^{11}$ hydrogen
nuclei per cm$^3$.

\item (iv) X-ray flares from similar star-disk encounters in the
nuclei of other LLAGN. On the one hand, the maximum X-ray luminosity
of flares at a given dimensionless distance $r_4$ from the black hole
is that given by equation \ref{lmax}, which behaves as $L_x\propto
n_{\rm ot}^2 \propto H^{-2}\propto 1/M_{\rm BH}^2$. That is more
massive black holes produce {\em weaker} X-ray flares. Further, the
background quiescent X-ray luminosity of confirmed LLAGN is always
larger than that of Sgr~A$^*$, usually by some 3-6 orders of
magnitude. On the other hand, some nearby Galaxies, e.g. M31 is not
detected in X-rays, whereas M32 has just been detected with $L_x\sim
10^{36}$ erg/sec (Ho et al. 2003). Hence flares in the nearby galaxies
could be detectable if similar inactive disks exist there. Clearly the
prospects of such discoveries in the future, in particular with XEUS
and Constellation-X missions, are much greater than they are with
Chandra now.

\item (v) X-ray spectra are expected to have a distribution of
properties. The brightest flares however should be produced by large
stars striking the disk at relatively small radii, and these should
have spectra as hard as $\Gamma\simeq 1.5$ with little \fe line
emission. Less luminous flares may have flatter spectra, $\Gamma\simgt
2$, and stronger \fe line emission.

\item (vi) Near infrared (NIR) luminosity of flares produced by the shock
around the star is around or smaller than the current detection limit
at 2.2$\mu$ (see \S \ref{sec:nir}).

\item (vii) No flare radio emission is expected in our model during
the star-disk impact. Such emission is completely self-absorbed in our
model.
\end{itemize}

\section{Shortcomings and limitations}\label{sec:short}

This paper presented a rather simplified treatment of the
problem. While aware of many of the potential complications or
modifications to our results, we kept the treatment at the present
level of complexity to have a relatively simple but broad view of the
different sides of the model. Nevertheless it is important to point
out some of these effects that we plan to study in the future.

First of all, as mentioned in \S \ref{sec:setup}, the exact geometry
of the star-disk passage is crucial for quantitatively accurate
results. Namely, the angle that the star makes with the disk rotation
velocity, $\theta_r$, controls the total mechanical energy deposition
rate. Here we simply assumed $\theta_r=\pi/2$. Further, the angle that
the star makes with the normal to the disk determines the velocity
component perpendicular to the disk, which is important for the
duration of the X-ray burst (e.g. see equation \ref{tdur}). In
addition, the relative position of the observer to the disk and the
star's direction of motion is also relevant. The emission of the
shocked gas viewed from the back side of the star is different than
that from the front side of the star.

Next complication is due to the fact that stars are of course not on
circular Keplerian orbits. In fact many of the observed stars have
highly elliptical orbits, many with $e\simeq 0.9$ (e.g. Genzel et
al. 2003; Ghez et al. 2003b). Therefore one needs to also allow for
variation in the absolute value of $v_*$ instead of simply using
$v_*=v_K$ as we have done in this paper. Similarly, we did not
consider the case of a very bright star such as S2 (e.g. \schodel et
al. 2002), or stars that produce substantial winds. Another complexity
is that the abundance of the putative disk in \sgra need not be
Solar. In fact if the disk is currently fed by the hot Helium rich
stars, then the Hydrogen in the disk may be under-abundant.
Additionally, \sgra star cluster may house a number of compact objects
-- white dwarfs, neutron stars and stellar mass black holes. These
would produce flares quite different from those that we have studied
here.

The aforementioned effects clearly demonstrate that our results may be
hoped to be representative of only the most frequent cases. Due to
these effects the spread in the properties of the expected flares
should be expected to be quite significant. These issues should be
addressed in the future with a more detailed likely numerical model.

\section{Discussion}\label{sec:discussion}

In this paper we proposed that the observed X-ray flares in \sgra are
emitted by bow shocks around stars when the latter pass through an
inactive disk. Such a disk has recently been invoked by Nayakshin
(2003) to explain the quiescent spectrum of Sgr A$^*$. The disk may be
a remnant of the past accretion and star formation activity in our
Galactic Center (\S \ref{sec:isdisk}; see also Levin \& Beloborodov
2003). Due to quiescent \sgra spectral constraints, the disk
temperature is restricted to be of order $10^2$ K, whereas the size of
the disk is at least few $\times 10^4$ gravitational radii. The disk
should be optically thin in mid infrared frequencies or else it would
violate the existing limits on the quiescent \sgra luminosity (see \S
\ref{sec:temperature}; Cuadra et al. 2003). From modeling X-ray flare
luminosities we concluded that the disk midplane density is of order
$10^{11}$ Hydrogen nuclei per cm$^3$. Such a disk is quite ``light''
-- it is not a subject to gravitational instability and it will not
gravitationally influence the dynamics of \sgra stellar cluster
significantly (see \S \ref{sec:disk}).

Using the latest results on star content in \sgra stellar cluster
(Genzel et al. 2003), we calculated the rate of the star-disk
crossings to be few per day (\S \ref{sec:cluster}). We then found that
the rate of star-disk crossings in the inner disk region with size
$R_i \sim 10^3 R_g$ is so large that the disk is very likely to be
completely emptied out (e.g. accreted) by the star passages within
astrophysically reasonable time of $\sim 10^6$ years.

In \S \ref{sec:single} we have shown schematics of a single star-disk
passage, and pointed out the relation of our work to the previous
extensive literature on the star-disk passages.  We then studied
properties of the expected X-ray flares in \S \ref{sec:xlum}, first in
the optically thin regime (\S \ref{sec:othin}), and then in the regime
in which the disk is optically thick to X-rays. In the optically thin
disk case most of the emission during the star-disk passage occurs in
the X-ray band. In the optically thick case, on the other hand, no
X-rays are observable from the shock front while it is inside the
disk. The emission is then reprocessed in the infrared. X-ray flares
are nevertheless are emitted when the shock arrives at the disk
photosphere. The maximum X-ray luminosity is emitted when the shock
front is at $\tau\sim 1$ where $\tau$ is effective optical depth to
X-rays along the line of sight. This then sets a very important limit
on the maximum X-ray luminosity that could be emitted in the star-disk
passages (equation \ref{lmax}). 

The predicted spectra range from very hard (bremsstrahlung dominated)
spectra with spectral index $\Gamma \sim 1.5$ to softer ones for
weaker X-ray flares with $\Gamma \simgt 2$ (\S \ref{sec:sot}).  The
hardest spectra are also predicted to show little \fe line emission
whereas softer flares should generally show the Helium and Hydrogen
like components of such lines. Finally, these lines could be
significantly red- or blue-shifted depending on whether the flare
occurs on the approaching or the the receding side of the disk:
$\Delta E_{\rm line} \sim (v_K/c)\;\sin{i}\; 6.7\, {\rm keV} = 50
\,\sin{i}\; {\rm eV} r_4^{-1/2}$. This in principle can be observable
with XMM and in future with Astro-E(2).

Comparing this maximum luminosity to the observed X-ray flare
luminosities we concluded that the disk must be somewhat optically
thin to X-rays at least in its innermost region. This then explains
why there is no excess absorption in the cold disk along the line of
site in our model, in accordance with the observations (Baganoff et
al. 2001, 2003b; Goldwurm et al. 2003a). In addition, we have shown
that the disk unshocked material is also photo-ionized by X-rays from
the shock (\S \ref{sec:photo}). For brightest ($L_x\simgt 10^{35}$
erg/sec) flares the photo-ionization alone could reduce the absorption
in the unshocked gas to the observed levels. Nevertheless weaker
flares, generally occurring at larger radii, may show signatures of
cold gas absorption as well as ``hot'' \fe lines.

Another important aspect of the predicted maximum X-ray flare
luminosity (equation \ref{lmax}) is that flares occurring at radii
greater than $\sim \hbox{few}\times 10^4 R_g$ are too weak to be
distinguishable on the background X-ray emission of Sgr~A$^*$. This
fact then explains why the flares have durations in the range from one
to tens of kiloseconds (see \S \ref{sec:typ}).

One of the most promising ways to confirm or rule out our model is
through observations of \fe line emission and/or absorption features
in X-ray spectra of flares. Discovery of such atomic features in \sgra
spectra would provide a very strong support for our model since these
atomic features are unlikely to be produced in the context of the
three other existing models for \sgra flares
(\citeNP{Markoffetal2001}; Liu \& Melia 2002; Narayan 2002, \S 5.3).
In all of these models flares come from a region very close to the
black hole, i.e., $\sim 10 R_g$, because otherwise the time scales of
the flares would be too long (Baganoff et al. 2001). The gas
(electron's) temperature is very large in this inner region, i.e., $T
\simgt 10^{10}$ K in the model by Narayan (2002, Fig. 5), $T\sim 10^9
-10^{11}$ K in Liu \& Melia (2002), and finally in the jet model of
Markoff et al. 2001 the gas is relativistic. These temperatures are
far too high to allow any noticeable \fe line emission. In addition,
gas densities are very low in some of these models (e.g., in an ADAF
or a CDAF), and the recombination time scale for completely ionized Fe
becomes many orders of magnitude longer than the burst duration, again
making \fe line emission implausible.

We also found that similar X-ray flares resulting from star-disk
passages in other LLAGN and nearby inactive Galaxies are rather hard
to detect except for the most nearby and {\em low black hole mass}
sources. As mentioned in \S \ref{sec:comp}, the predicted maximum
X-ray flare luminosity (equation \ref{lmax}) behaves as $\lmax\propto
1/M_{\rm BH}^2$. In contrast, the quiescent X-ray luminosity is
proportional to $\propto \dot{m} M_{\rm BH}$ where $\dot{m}$ is the
dimensionless accretion rate (in the usual Eddington units). Hence it
becomes more and more difficult to discover such flares with increase
in the black hole mass. Yet some of the nearby sources may show
detectable flares. For example, the recently detected nuclear X-ray
source in M32 has $\sim 0.5-5$ keV luminosity of about $10^{36}$
erg/sec (Ho et al. 2003; the black hole mass in M32 is very similar to
that of Sgr A$^*$). This luminosity is low enough to allow detection
of the strongest X-ray flares, especially in the future with
instruments such as XEUS and Constellation-X. Galaxies not yet
detected in X-rays, such as M31, hold an even greater promise for
future detection of X-ray flares.

Another very promising way to confirm or refute our model is through
detection of a NIR flare counterpart. In our model these flares are
emitted some $\sim 10^3-10^4 R_g$ off \sgra location, which translates
to $\sim 0.01 -0.1$ arcseconds. Current observational techniques do
allow to measure such a small angular offset to be measured (Genzel et
al. 2003; Ghez et al. 2003b). Unfortunately the NIR emission from
flares was predicted to be relatively weak (\S \ref{sec:nir}),
although more work is needed to refine this.  The properties of the
disk can also be constrained via stellar eclipses and other similar
effects (Nayakshin \& Sunyaev 2003; Cuadra et al. 2003).

\del{We conclude by noting that our model does remarkably well in
explaining the current observational status of \sgra flares. We
therefore feel strongly that {\em a cold disk of some sort} must be
present in the central $\sim 0.1-1$ arcsecond (or $10^4 -10^5 R_g$, or
$ 10^{16}-10^{17}$ cm) of our Galaxy. This would further strengthen
the case for the physical similarity between \sgra and the quiescent
centers of other galaxies, and LLAGN in particular, which are known to
have such cold disks (e.g. Miyoshi et al. 1995; Quataert et al. 1999;
Ho 2003). We also believe that implications of the disk presence may
be very significant for the theory of accretion flows at low accretion
rates (Nayakshin 2003).}

\section*{ACKNOWLEDGMENTS}

We are very thankful for receiving the drafts of two not yet published
papers from R. Genzel, R. Sch\"odel and T. Ott, and for useful
discussions that followed. We also appreciate discussions with
P. Predehl, A. Goldwurm \& A. Ghez. Comments from many of our
colleagues at MPA in Garching have also been very useful.


{}

\end{document}